\documentclass[aps,prx,reprint,superscriptaddress]{revtex4-2}
\usepackage{amsmath}
\usepackage{amsfonts}
\usepackage{amssymb}
\usepackage{graphicx}
\usepackage{booktabs}
\usepackage[colorlinks,linkcolor=blue,anchorcolor=blue,citecolor=blue,urlcolor=blue]{hyperref}
\usepackage{booktabs}
\usepackage{makecell}
\usepackage{diagbox}
\usepackage{colortbl}
\usepackage{xcolor}
\usepackage{adjustbox}

\usepackage{mathrsfs}
\usepackage{epsfig}
\usepackage{float}
\usepackage{subfigure}
\usepackage{mathtools} 
\usepackage{color}

\usepackage{ulem}
\newcommand{\stkout}[1]{\ifmmode\text{\sout{\ensuremath{#1}}}\else\sout{#1}\fi}
\begin{document}
\title{Stochastic Mean-field Theory for Conditional Spin Squeezing by Homodyne Probing of Atoms-Photon Dressed States}

\author{Zhiqing Zhang}
\affiliation{Henan Key Laboratory of Diamond Optoelectronic Materials and Devices, Key Laboratory of Material Physics Ministry of Education, School of Physics and Laboratory of Zhongyuan Light, Zhengzhou University, Zhengzhou 450052, China}

\author{HaiZhong Guo}
\affiliation{Henan Key Laboratory of Diamond Optoelectronic Materials and Devices, Key Laboratory of Material Physics Ministry of Education, School of Physics and Laboratory of Zhongyuan Light, Zhengzhou University, Zhengzhou 450052, China}
\affiliation{Institute of Quantum Materials and Physics, Henan Academy of Sciences, Zhengzhou 450046, China}

\author{Lingrui Wang}
\affiliation{Henan Key Laboratory of Diamond Optoelectronic Materials and Devices, Key Laboratory of Material Physics Ministry of Education, School of Physics and Laboratory of Zhongyuan Light, Zhengzhou University, Zhengzhou 450052, China}

\author{Gang Chen}
\affiliation{Henan Key Laboratory of Diamond Optoelectronic Materials and Devices, Key Laboratory of Material Physics Ministry of Education, School of Physics and Laboratory of Zhongyuan Light, Zhengzhou University, Zhengzhou 450052, China}
\affiliation{Institute of Quantum Materials and Physics, Henan Academy of Sciences, Zhengzhou 450046, China}

\author{Chongxin Shan}
\email{cxshan@zzu.edu.cn}
\affiliation{Henan Key Laboratory of Diamond Optoelectronic Materials and Devices, Key Laboratory of Material Physics Ministry of Education, School of Physics and Laboratory of Zhongyuan Light, Zhengzhou University, Zhengzhou 450052, China}
\affiliation{Institute of Quantum Materials and Physics, Henan Academy of Sciences, Zhengzhou 450046, China}

\author{Klaus M{\o}lmer}
\email{klaus.molmer@nbi.ku.dk}
\affiliation{Niels Bohr Institute, University of Copenhagen, 2100 Copenhagen, Denmark}

\author{Yuan Zhang}
\email{yzhuaudipc@zzu.edu.cn}
\affiliation{Henan Key Laboratory of Diamond Optoelectronic Materials and Devices, Key Laboratory of Material Physics Ministry of Education, School of Physics and Laboratory of Zhongyuan Light, Zhengzhou University, Zhengzhou 450052, China}
\affiliation{Institute of Quantum Materials and Physics, Henan Academy of Sciences, Zhengzhou 450046, China}

\begin{abstract}
Projective measurements of collective observables can be employed to herald the preparation of entangled states of quantum systems, and the resulting conditional dynamics is usually handled by stochastic master equation (SME) for small systems, and by an approximate Gaussian-state formalism for large systems. In this work, we present an alternative technique by developing a stochastic variant of cumulant mean-field theory, benchmark it against an exact stochastic collective density matrix approach by the simulations of hundreds of identical two-level atoms. More importantly, we demonstrate its full power by studying the conditional spin squeezing of thousands of three-level atoms coupled strongly with an optical cavity subject to individual decay and dephasing, and by simulating the experimental protocol to reveal formation and detection of the spin squeezed state. The proposed technique might be further extended to study more exotic quantum-measurement effects of large quantum systems, such as deterministic spin squeezing with quantum feedback, spin squeezing of optical clock transitions, and retrodictive spin squeezing by posterior measurements, and so on. 
\end{abstract}
\maketitle

\section{Introduction}
Spin squeezing in atomic ensembles has attracted considerable interest due to its important applications in quantum information science~\cite{BJulsgaard,BJulsgaard2004,LPezz2021,YSha2023,BKM2022} and high-precision metrology~\cite{DJWineland,VMeyer,ALChauvet,EPP2020,GPG2022,JMR2024,JD2025}, and the witness of multipartite entanglement~\cite{OGuhne,AS2001,JE2008,GT2009,LP2018}. Such nonclassical states can be generated through coupling to squeezed light~\cite{AK1997,JHald}, collisional interactions~\cite{GPG2022,CG2010,IDL2010,MAN2018}, and quantum non-demolition (QND) measurements~\cite{GPG2022,AKuzmich,ZChen,KCCox,MHSchleier,OHosten,ZYLi}.
In the QND measurements, the probed observables commute with the system Hamiltonian and are not altered by the measurements \cite{VBBraginsky}, while the accumulated knowledge yields the conditional squeezing.  Conditional spin squeezing can be realized by detecting optical phase shift or frequency shift induced by collective populations of bare atomic states \cite{MHSchleier,OHosten,IBouchoule,AEBNielsen}, or atoms-photon dressed states \cite{ZChen,JGBohnet}. By applying feedback to the system, it is also possible to compensate the effect of the random measurement outcomes, and to achieve deterministic spin squeezing \cite{LKThomsen,LKThomsenPRA,KCCox}.

Within the quantum measurement theory, stochastic master equation (SME) describes the quantum systems subject to continuous measurements \cite{HMWiseman}. The standard technique to solve the SME is the density matrix technique, but it is normally limited to the systems with few atoms due to the exponentially increased Hilbert space~\cite{JRFohansson}. By exploring permutation symmetry of identical atoms, the density matrix approach based on collective numbers or Dicke states can be applied to simulate systems with hundreds of atoms~\cite{MACRossi2020,YZhang}. By characterizing the atoms with means and covariances of collective operators, the Gaussian state formulism permits the simulations of systems with unlimited number of atoms~\cite{BLMadsen}, but has the difficulty of incorporating properly the individual atomic dissipation and handling multi-level systems. 

In this work, we present an alternative technique by developing a stochastic variant of the cumulant mean-field theory~\cite{DPlankensteiner}, which can handle many aspects of the realistic systems, such as tens of thousands of atoms, multiple atomic levels, dissipation of individual atoms, strong atoms-photon coupling, and inhomogeneous coupling to the cavity and so on. A more detailed comparison between the proposed approach and the existing methods is provided in Appendix~\ref{sec:compare} and Ref.~\citep{YZhang}. Furthermore, we have implemented the proposed approach by combining QuantumCumulnats.jl and StochasticDiffEq.jl package, which permits automatic deriving and solving of stochastic mean-field equations, and thus lowers the barrier of applying the proposed approach to different problems. 

In the following, we benchmark the proposed approach against an exact approach by solving a standard SME for hundreds of two-level atoms subject to a QND measurement, obtained by adiabatic eliminating higher atomic excited levels in Raman scattering processes~\citep{HMWiseman}. Then, we demonstrate the power of the approach by developing an extended SME for tens of thousands of three-level atoms coupled strongly with an optical cavity subject to a QND detection, where the formed atoms-photon dressed states prevents the adiabatic elimination, by simulating the experiential spin squeezing protocol with four laser pulses to provide insights into the involved dynamics, which shows a remarkable consistency with earlier experiments ~\cite{ZChen, JGBohnet}. In the future, the proposed approach can be further developed to explore more exotic quantum-measurement effects of large quantum systems ~\cite{EPP2020,JMR2024,LKThomsen,LKThomsenPRA,KCCox,HBao,HBaoNC,AShankar}. 

\section{Benchmark of the Proposed Approach with A Standard System} 
To benchmark the proposed approach in the context of conditional spin squeezing, we consider firstly the standard SME for density operator $\hat{\rho}_I$ of a two-level atomic ensemble conditioned on the QND measurement outcome $I$ \cite{LKThomsen}: 
\begin{equation}
\partial_t \hat{\rho}_I = -M  \mathcal{D}[\hat{J}_z]\hat{\rho}_I  + \sqrt{M} (dW/dt) \mathcal{H}[\hat{J}_z]\hat{\rho}_I, \label{eq:standard-sme}
\end{equation}
where the two terms on the right side describe the collective dephasing and the QND measurement backaction with a strength $M$. The super-operators are defined as  $\mathcal{D}[\hat{o}]\hat{\rho}_I = \left(\hat{o}^{\dagger}\hat{o}\hat{\rho}_I+\hat{\rho}_I \hat{o}^{\dagger}\hat{o}\right)/2-\hat{o}\hat{\rho}_I \hat{o}^{\dagger}$  and $\mathcal{H}[\hat{o}]\hat{\rho}_I = \hat{o}\hat{\rho}_I + \hat{\rho}_I\hat{o}^\dagger - (\langle \hat{o}\rangle + \langle \hat{o}^\dagger\rangle)  \hat{\rho}_I$ for any operator $\hat{o}$ and its expectation value $\langle \hat{o}\rangle = \mathrm{tr}\left\{ \hat{o} \hat{\rho}_I\right\}$. To characterize the states of the atomic ensemble, we introduce a collective spin vector operator $\hat{J}= \sum_{i=x,y,z} \hat{J}_i e_i$ with the components $\hat{J}_i=\sum_{k=1}^N\hat{\sigma}_k^i$, and compute the mean values of the component operators $J_i=\langle \hat{J}_i \rangle$ and their uncertainties $\Delta J_i=\sqrt{\langle \hat{J}_i^2 \rangle-\langle \hat{J}_i \rangle^2}$, from which the spin squeezing parameters~\cite{JMa} $\xi_{i}^{2}= N\left(\Delta J_{i}\right)^{2}/(J_{j}^{2}+J_{k}^{2})$ (with $i,j,k$ are the cycling of $x,y,z$) can be computed. Here, $\hat{\sigma}_{k}^i$ are the three Pauli operators of the $k$-th atom of the total $N$  atoms. The random numbers $dW$ account for photon white-noise of the homodyne or heterodyne detection, and follow a normal distribution with zero mean and variance $dt$, i.e. $\mathrm{E}\left(dW\right)=0$ and $dW^2=dt$. 

In our proposed approach, we solve the SME~(\ref{eq:standard-sme}) by deriving equations of motion $\partial_t \left\langle \hat{o} \right\rangle = {\rm tr}\{(\partial_t \hat{\rho}_I) \hat{o}\}$ for the expectation value $\left\langle \hat{o} \right\rangle$ of generic operator $\hat{o}$, and  obtain the equations for the mean values of single atom $\left \langle \hat{\sigma}_{k}^i \right \rangle$, which couple to the atom-atom correlations $\langle \hat{\sigma}_{k}^i \hat{\sigma}_{k'}^j  \rangle$ (for $k\neq k'$). The correlations in turn couple to the mean values of more operators, leading to a hierarchy of equations, and we employ second-order cumulant expansion approximation $\langle\hat{o}\hat{p}\hat{q}\rangle=\left \langle \hat{o}\right \rangle \left \langle \hat{p}\hat{q} \right \rangle + \left \langle \hat{p}\right \rangle \left \langle \hat{o}\hat{q} \right \rangle + \left \langle \hat{q}\right \rangle \left \langle \hat{o}\hat{p} \right \rangle - 2 \left \langle \hat{o}\right \rangle \left \langle \hat{p} \right \rangle \left \langle \hat{q} \right \rangle$ for single-particle and single-photon operators $\hat{o},\hat{p},\hat{q}$, to truncate the hierarchy. Furthermore, by assuming that all the atoms are identical, $\left \langle \hat{\sigma}_{k}^i \right \rangle$ are the same for all the atoms, and $\langle \hat{\sigma}_{k}^i \hat{\sigma}_{k'}^{j} \rangle$ are same for all atom pairs $(k,k')$ \cite{DMeiser2009}. As a result, the number of independent equations reduces from the order of $\sim N^2$ to only a few tens, and the number of atoms $N$ enters the formalism only as a parameter, enabling efficient simulations even for large ensembles. In the Appendix \ref{subsec:mean}, we present the derived  stochastic mean-field equations and the formulas to compute the spin squeezing parameters.

\begin{figure}[tbph]
\centering
\includegraphics[scale=0.5]{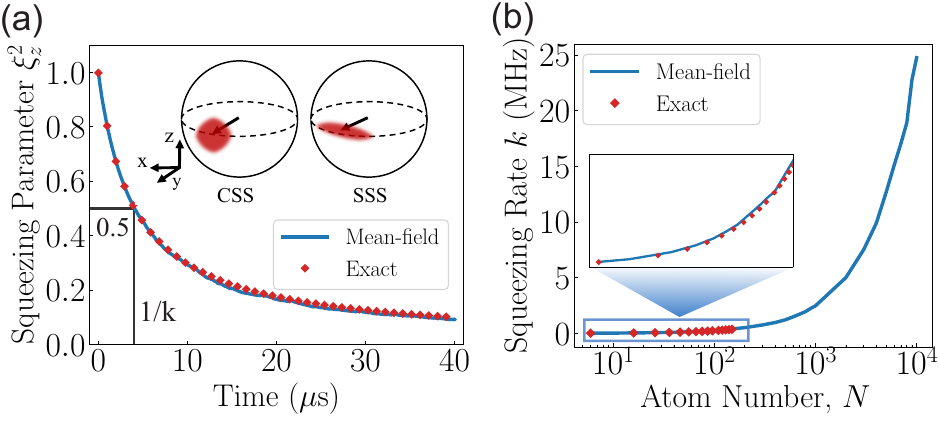}
\caption{\label{fig:benchmark}
 Benchmark of the results based on the proposed stochastic mean-field approach (solid line) and the exact stochastic collective density matrix approach (dots). (a) shows the spin squeezing parameter $\xi_z^2$ for a system with $N=100$ atoms evolving from a coherent spin state (CSS, left inset) to the spin squeezed state (SSS, right inset), which can be well fitted with a function~\citep{BLMadsen} $\xi_z^2(t)=1/(1+kt)$ with a spin squeezing rate $k$. (b) shows $k$ as function of the number of atoms $N$.}
\end{figure}

To benchmark the proposed approach, we employ an exact numerical approach based on collective density matrix~\citep{YZhang} to solve the standard SME~\eqref{eq:standard-sme}. In this approach, we assume that all the atoms are identical, and map the density matrix elements to single quantity, whose number scales only cubically with the number of atoms $\sim N^3$. The details of such an approach can be found in our previous article~\cite{YZhang} and Appendix~\ref{subsec:coldm}. Figure~\ref{fig:benchmark}(a) displays the time evolution of the spin squeezing parameter $\xi_{z}^{2}$ for $N=100$ atoms, initially prepared in a coherent spin state (CSS) with the collective spin vector pointing along the y-axis (left inset)  and evolving into a spin-squeezed state (SSS, right inset). The results from our stochastic mean-field approach (solid blue line) show excellent agreement with the exact simulations (red dots). The decay of $\xi_{z}^{2}(t)$  is well characterized by the functional form  $\xi_{z}^{2}(t)=1/(1+kt)$ derived in Ref.~\citep{BLMadsen}, which defines a spin squeezing rate $k$. We extract this rate $k$ and plot it as a function of atom number $N$ in Fig.~\ref{fig:benchmark}(b). The two methods remain in remarkable agreement for systems up to $N=150$ atoms.  This benchmark validates our approximate approach and demonstrates its applicability to systems exceeding $10^4$ atoms, as extrapolated in [Fig.~\ref{fig:benchmark}(b)]. 

\section{Application of the Proposed Approach to A Realistic System}
\begin{figure}
\begin{centering}
\includegraphics[scale=0.4]{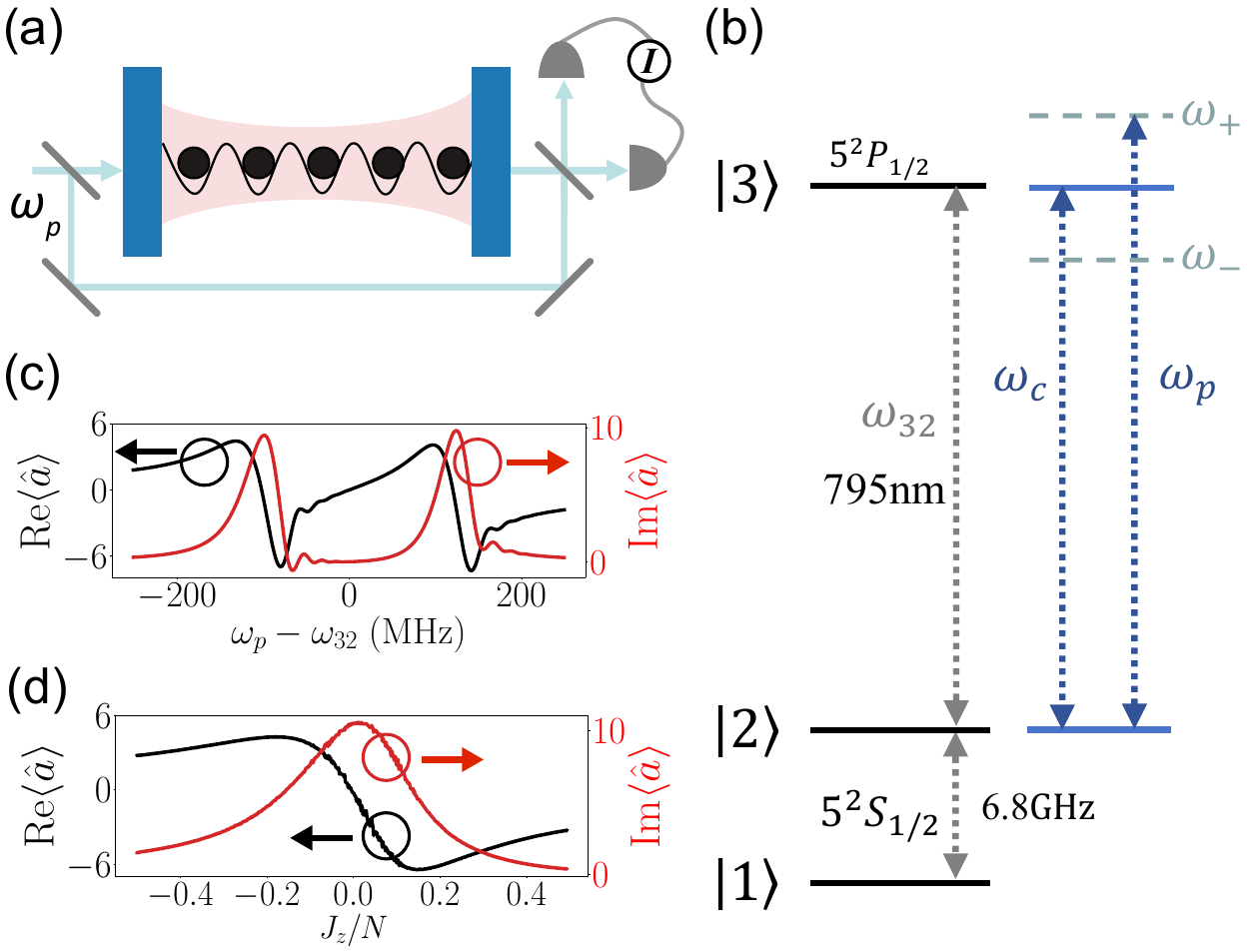}
\par\end{centering}
\caption{System and energy diagram. (a) shows $N$ rubidium-87 atoms inside an optical cavity, and the balanced homodyne detection of the field formed by mixing the probe field at frequency $\omega_{p}$ with the field transmitted through the cavity. (b) shows the simplified energy diagram of the atoms with two hyper-fine ground states, represented as the up $|2\rangle$ and down $|1\rangle$ state of a spin-1/2 particle, and an electronic excited state $|3\rangle$, coupling strongly with the optical cavity of frequency $\omega_c$ , which leads to two atoms-photon dressed states of transition frequencies $\omega_{\pm}$ (horizontal dashed lines). (c,d) show the intra-cavity field amplitude $\langle \hat{a} \rangle$ as function of $\omega_p$ relative to the atomic transition frequency $\omega_{32}$ (c) for the atomic ensemble prepared in an equal superposition of the two hyper-fine ground states, and the imbalance of the two ground state populations $J_z/N$ for $\omega_p =\omega_{32} + \sqrt{N/2}g$ (d). For more details see text.} 
\label{fig:SqueezingQND}
\end{figure}

We explore now the full power of the proposed approach by examining a realistic system, where the conditional spin squeezing is realized by a homodyne probing of $N=10^4$ rubidium-87 atoms in an optical cavity with a probe laser of frequency $\omega_p$ [Fig.~\ref{fig:SqueezingQND}(a)]. Note that the analysis presented below applies also to the earlier experiments \cite{ZChen,JGBohnet} employing a heterodyne detection scheme. In our system, we treat two atomic hyper-fine ground states, $|1\rangle=\left|5^{2}\mathrm{S}_{1/2},F=2,m_{F}=0\right\rangle$ and $|2\rangle=\left|5^{2}\mathrm{S}_{1/2},F=1,m_{F}=0\right\rangle$, as the up and down states of a pseudo spin-$1/2$ particle. An electronic excited state $|3\rangle=\left|5^{2}\mathrm{P}_{1/2},F'=1,m_{F}=0\right\rangle$ couples strongly with  an optical cavity mode with strength $g$ (single-atom coupling strength), leading to two atoms-photon dressed states  [Fig.~\ref{fig:SqueezingQND}(b)]. In the homodyne detection, the photo-current is proportional to the real part of the intra-cavity field amplitude ${\rm Re}\langle \hat{a}\rangle$. This amplitude explores the resonances around the dressed states energies [Fig.~\ref{fig:SqueezingQND}(c)], which are associated with the population on the upper hyper-fine ground state, $J_z +N/2$ [Fig.~\ref{fig:SqueezingQND}(d)].  The homodyne signal thus measures the collective spin z-component $J_z$, and squeezes its uncertainty $\Delta J_z$.

To describe the conditional dynamics of the system, we introduce the extended SME:
\begin{align}
 & \frac{\partial}{\partial t}\hat{\rho}_{I}=-\frac{i}{\hbar}\left[\hat{H}_{c}+\hat{H}_{p}+\hat{H}_{a}+\hat{H}_{a-c}+\hat{H}_{a-m},\hat{\rho}_{I}\right]\nonumber \\
 & -\kappa\mathcal{D}\left[\hat{a}\right]\hat{\rho}_{I}-\gamma\sum_{k=1}^N\mathcal{D}\left[\hat{\sigma}_{k}^{23}\right]\hat{\rho}_{I}-\frac{\chi}{2} \sum_{k=1}^N\mathcal{D}\left[\hat{\sigma}_{k}^{22} - \hat{\sigma}_{k}^{33}\right]\hat{\rho}_{I}\nonumber \\
 & +\frac{dW}{dt}\sqrt{\eta\kappa_{2}}\left[e^{i \omega_p t}\left(\hat{a}-\left\langle \hat{a}\right\rangle \right)\hat{\rho}_{I}+e^{-i \omega_p t}\hat{\rho}_{I}\left(\hat{a}^{\dagger}-\left\langle \hat{a}^{\dagger}\right\rangle \right)\right].\label{eq:sme}
\end{align}
The optical cavity is described by the Hamiltonian $\hat{H}_{c}=\hbar\omega_{c}\hat{a}^{\dagger}\hat{a}$ with the frequency $\omega_{c}$, the photon creation operator $\hat{a}^{\dagger}$ and annihilation operator $\hat{a}$. The probing of the cavity by a laser beam of frequency $\omega_{p}$ is described by the Hamiltonian $\hat{H}_{p}=\hbar\Omega_p\sqrt{\kappa_{1}}\left(e^{i\omega_{p}t}\hat{a}+e^{-i\omega_{p}t}\hat{a}^{\dagger}\right)$ with the transmission coefficient $\sqrt{\kappa_{1}}$ ($\kappa_{1},\kappa_{2}$ are the photon loss rates due to the left and right mirrors), and the probe amplitude parameterized by the strength $\Omega_p$. The  $N$ atoms ensemble is described by the Hamiltonian $\hat{H}_{a}=\hbar\sum_{k=1}^{N}\left(\omega_{21}\hat{\sigma}_{k}^{22}+\omega_{31}\hat{\sigma}_{k}^{33}\right)$ with the transition frequencies $\omega_{21},\omega_{31}$ [Fig.~\ref{fig:SqueezingQND}(b)]. Here and in the following, we define the atomic  operators as $\hat{\sigma}_{k}^{ll'}=\left|l_{k}\right\rangle \left\langle l'_{k}\right|$ with $l,l'=1,2,3$. The atoms-cavity interaction is described by the Hamiltonian $\hat{H}_{a-c}=\hbar g\sum_{k=1}^N\left(\hat{a}^{\dagger}\hat{\sigma}_{k}^{23}+\hat{\sigma}_{k}^{32}\hat{a}\right)$ with the coupling strength $g$. In addition, a classical microwave field  with frequency $\omega_{m}$ is introduced to manipulate the states of the pseudo spins via the Hamiltonian $\hat{H}_{a-m}=\hbar \Omega_m \sum_{k=1}^N \left(e^{i\omega_{m}t}\hat{\sigma}_{k}^{12}+e^{-i\omega_{m}t} \hat{\sigma}_{k}^{21}\right)$ with strength $\Omega_m$.

The second line of Eq. (\ref{eq:sme}) describes the system dissipation, which includes the cavity photon loss with a rate $\kappa=\kappa_{1}+\kappa_{2}$,  the spontaneous emission and dephasing of individual atoms (associated with the upper hyper-fine ground state) with rates $\gamma$, $\chi$, respectively. The last line of Eq. (\ref{eq:sme}) describes the backaction of the homodyne detection with the photon-shot noise, modeled by a Wiener increment $dW$. The photocurrent difference between the two photodetectors (with an efficiency $\eta$)  $I\left(t\right)=\sqrt{\eta\kappa_{2}}\mathrm{Re}\left\langle \hat{a}\right\rangle(t) +dW/dt$ is proportional to the real part of the intra-cavity field amplitude ${\rm Re} \left\langle \hat{a}\right\rangle$ but is dominated by the white noise $dW/dt$ at short time. 

Because of the strong atoms-photon coupling, we can not eliminate the atomic excited level, and thus have to solve Eq.~\eqref{eq:sme} explicitly. To avoid the errors in equations derivation, we have generalized the QuantumCumulants.jl package, developed by Helmut Ritsch's group to solve deterministic master equations with the mean-field approach, to automatically derive and solve the stochastic mean-field equations in our approach. In the Appendix~\ref{sec:code} and~\ref{sec:meaneq}, we present the corresponding codes and the derived equations. In the following, we focus on the numerical results. We have used the parameters compatible with the experiments in Ref.~\citep{ZChen} and Ref.~\citep{JGBohnet}, and summarize them in Appendix~\ref{sec:supresult}. There, the key parameters are the number of atoms $N=10^4$, the atoms-photon coupling $g=2\pi \times 253$ kHz, the photon damping rate $\kappa=2\pi \times 11.1$ MHz and the atomic spontaneous emission rate $\gamma=2\pi \times 5.75$ MHz.

\begin{figure}
\begin{centering}
\includegraphics[scale=0.30]
{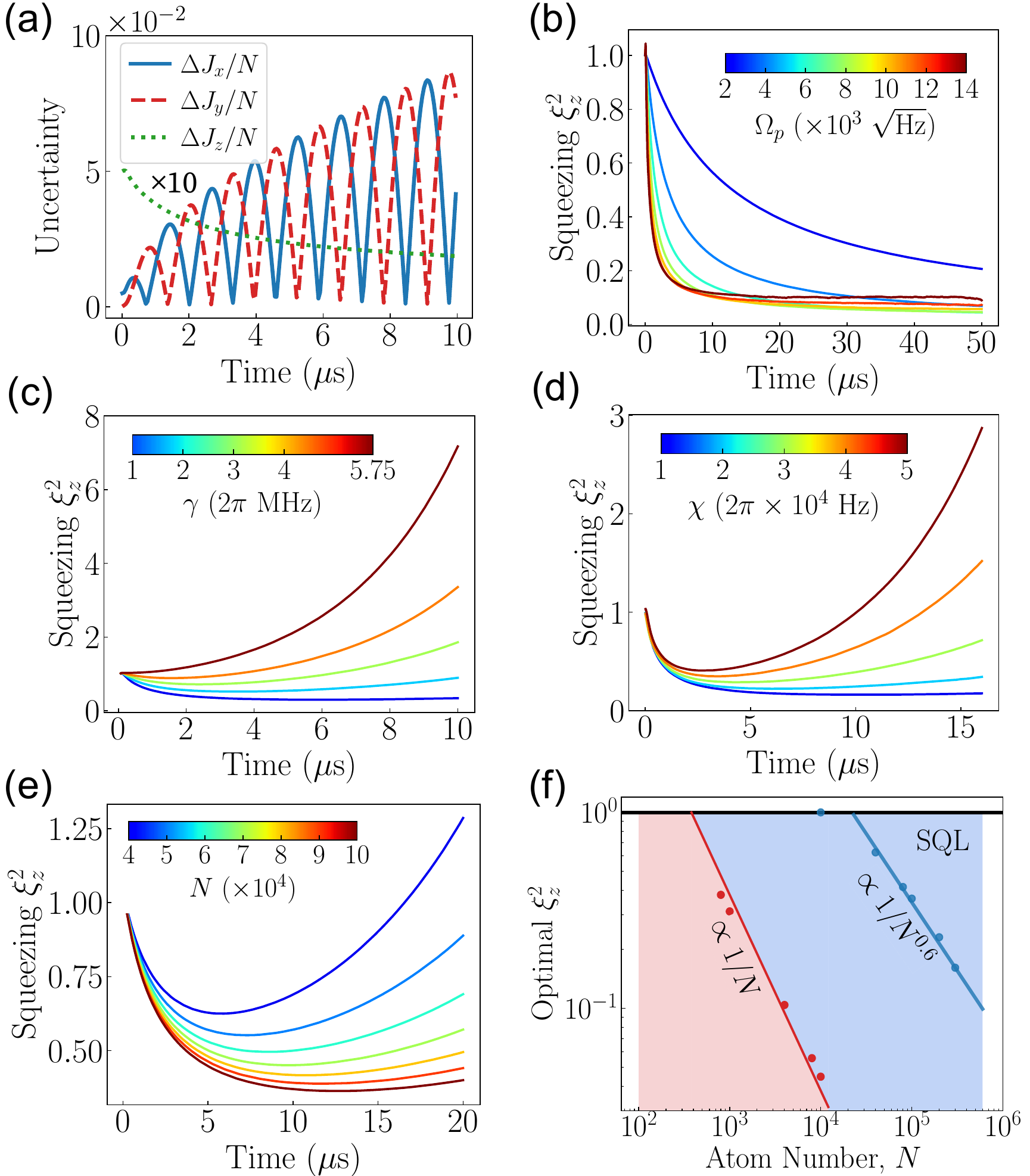}
\par\end{centering}
\caption{ Conditional spin squeezing dynamics. (a) shows the evolution of uncertainty of the collective spin components $\Delta J_x$,$\Delta J_y$,$\Delta J_z$ for an ideal system without the individual atomic dissipation, where $\Delta J_z$ is amplified tenfold for visual clarity, and all quantities are normalized to the atomic number $N$. (b-e) show the spin squeezing parameter $\xi_z^2$ as function of time for different probe amplitudes $\Omega_p$, atomic decay rates $\gamma$, atomic dephasing rates $\chi$ and numbers of atoms $N$, respectively. (f) shows the $\xi_z^2=1$ (the standard quantum limit, SQL) for the systems in the CSS (black line), and the reduction of the minimal $\xi_z^2$ with increasing $N$ for the systems in the absence (red dots and line) and presence (blue dots and line) of the individual atomic dissipations, which can be fitted with $\xi_z^2 \sim 1/N$ (i.e. Heisenberg limit, HL) and $\xi_z^2 \sim 1/N^{0.6}$ (between the SQL and the HL), respectively.}
\label{fig:con_spin_squeezing}
\end{figure}

In Fig.~\ref{fig:con_spin_squeezing}, we show the behavior of the uncertainty of collective spin components and the spin squeezing parameter during the homodyne detection. Here, we prepare the atomic ensemble in a CSS such that the collective spin points along the y-axis and thus has equal uncertainty along the x- and z-axis. Our calculations show that during the laser probing, the components $J_x$ and $J_y$ oscillate with time, while the component $J_z$ is slightly reduced. At the same time, the uncertainty $\Delta J_z$  decreases while  $\Delta J_x,\Delta J_y$ oscillate in time with increasing amplitude [Fig.~\ref{fig:con_spin_squeezing}(a)]. These results indicate that the collective spin vector $\mathbf{J}$ rotates around the z-axis, which can be explained by the probe field-induced AC Stack shift (Fig.~\ref{fig:rotation} of Appendix~\ref{sec:supresult}), and the spin squeezing and anti-squeezing occur for the projection of the collective spin vector along the z-axis and  in the equatorial plane, respectively. 

In Fig.~\ref{fig:con_spin_squeezing}(b), we study the dependence of the spin squeezing parameter $\xi_z^2(t)$ on the probe field without individual atomic dissipation. As the probe strength increases, the spin squeezing occurs faster and its minimal value increases slightly, which agree qualitatively with the experimental results \cite{JGBohnet}. In Fig.~\ref{fig:con_spin_squeezing}(c,d), we investigate how the decay $\gamma$  and dephasing $\chi$  of individual atoms affect the conditional spin squeezing.  In the presence of $\gamma$, the uncertainty $\Delta J_z$ decrease with time as in the ideal case, while $J_\parallel=\sqrt{J_x^2+J_y^2}$ also decrease. As a result, the individual decay affects the spin squeezing mainly by reducing the length of the collective spin vector, and the spin squeezing parameter $\xi_z^2$ increases [Fig.~\ref{fig:con_spin_squeezing}(c)]. The atomic dephasing rate $\chi$ affects the conditional spin squeezing in a similar way [Fig.~\ref{fig:con_spin_squeezing}(d)], except that the collective spin z-component $J_z$ also decreases with time  (not shown). In addition, we find that the squeezing parameter can be fitted with the expression \cite{BLMadsen} $\xi_z^2(t)=A/(1+k_1 t) + (1-A)e^{k_2 t}$, from which the pertaining squeezing rate $k_1$ and the anti-squeezing rate $k_2$ can be quantitatively determined. In Fig.~\ref{fig:detection} and Fig.~\ref{fig:verif} of Appendix~\ref{sec:supresult}, we have further examined the influence of the detection efficiency and the laser-atom detuning, as well as the fitted squeezing and anti-squeezing rate.

The detrimental effects of individual atomic dissipation can be partially mitigated by increasing the number of atoms $N$ [Fig.~\ref{fig:con_spin_squeezing}(e)]. As $N$ increases from $4\times 10^4$ to $10^5$, the initial decrease and the latter raising of the spin squeezing parameter $\xi_z^2$ becomes faster and slower, respectively. At the same time, the minimal $\xi_z^2$ also decreases and occurs later. In Fig.~\ref{fig:con_spin_squeezing}(f), we plot this value as a funtion of $N$. For the system in a CSS, we have $\xi_z^2=1$ (black line), indicating the standard quantum limit (SQL). For the ideal system without the atomic dissipation, the results depart from  the SQL line for $N$ larger than 500, and can be well fitted with the formula $\sim 1/N$ (red dots and line), hinting at the Heisenberg limit (HL). For the realistic system with the atomic dissipation, the results deviate from the SQL line for $N$ larger than $10^4$, and can be fitted with $\sim 1/N^{0.6}$ (blue dots and line), indicating a scaling between the SQL and the HL. Note that the blue dots agree qualitatively with the experimental results (cf. Fig.~4 of Ref.\cite{JGBohnet}). 

We have also investigated the influence of the frequency detuning $\Delta=\omega_{32}-\omega_c$ between the upper hyperfine ground-excited state transition $\omega_{32}$ and the cavity mode $\omega_c$, and presented the corresponding results in Appendix~\ref{sec:supresult}. The similar results are found for the probe field resonant to the lower dressed state, except that the AC Stark effect-induced collective spin vector rotation occurs for the opposite direction.

\section{Simulation of Experimental Protocol} 
\begin{figure}
\begin{centering}
\includegraphics[scale=0.29]{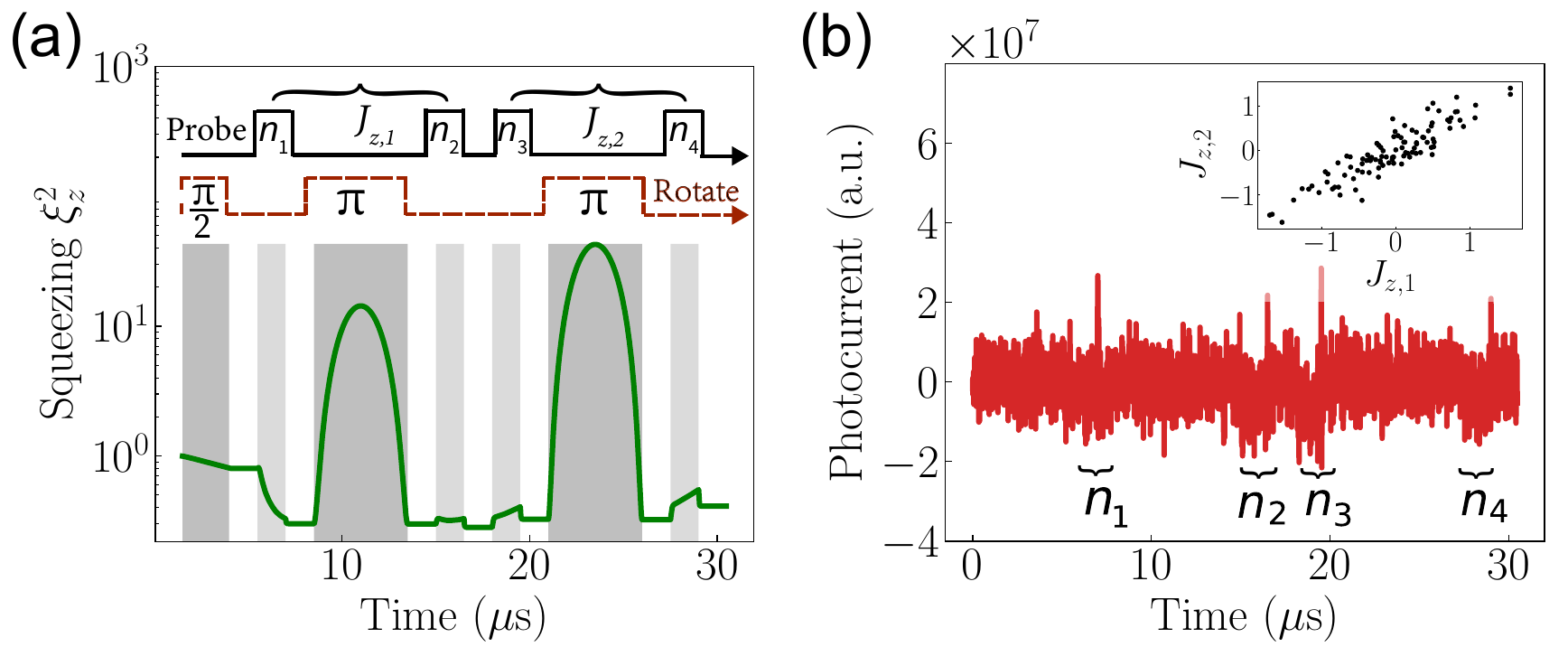}
\par\end{centering}
\caption{ \label{fig:sim} Simulation of experimental protocol. (a) shows the spin squeezing parameter $\xi_z^2$ following the protocol shown on the top. (b) shows the dynamic of the photo-current in the homodyne detection, where the current during the four laser pulses are integrated to form the quantities $n_i$ ($i=1,2,3,4$), and the inset shows $J_{z,2}=n_4 -n_3$ versus $J_{z,1}=n_1 -n_2$ during the preparation and verification of the conditional spin squeezing for one hundred simulations. For more details see the text.}
\label{fig:dep}
\end{figure}

Next, we apply the proposed approach to simulate the experimental spin squeezing protocol~\cite{ZChen} in order to achieve insights into the involved dynamics (Fig.~\ref{fig:sim}). In this protocol, a microwave $\pi/2$-pulse is firstly applied to prepare the atomic ensemble in a CSS, and then two laser probe pulses intersected by a microwave $\pi$-pulse are applied to make the atomic ensemble evolve into the  SSS, and finally the procedure in the second step is repeated to read out the  SSS [upper inset of  Fig.~\ref{fig:sim}(a)].   

The simulated photocurrent of the homodyne detection is shown in Fig.~\ref{fig:sim}(b). Following the protocol, we denote the integrated current as $n_i$ (with  $i=1,2,3,4$) during the application of the four laser  pulses, and utilize the differences $J_{z,1} = n_1 - n_2$, $J_{z,2} = n_4 - n_3$ as an estimation of the collective spin z-component $J_z$ for the prepared and probed SSS, respectively. Then, the conditional spin squeezing can be confirmed by a correlation between $J_{z,1}$ and $J_{z,2}$. Despite of the noisy photocurrent, we have indeed observed the expected correlation [inset of Fig.~\ref{fig:sim}(b)]. By associating with Fig.~\ref{fig:sim}(a), we perceive that the spin squeezing drops fast in the first probe pulse, and raises slightly for the remaining probe pulses, suggesting an averaging effect due to the finite laser pulses. Thus, the degree of the squeezing as inferred in the experiments might be smaller than what is actually achieved. In addition, despite of different detection scheme (homodyne versus heterodyne detection), we find that the observed correlation agrees qualitatively with that achieved in the experiment [see Fig.~\ref{fig:con_spin_squeezing}(a) of Ref. \cite{ZChen}].

\section{Conclusions} In summary, we have developed a stochastic mean-field approach to address the conditional spin squeezing of large quantum systems, benchmarked it against an exact approach based on collective density matrix, and demonstrated its power through the simulations of more complex but realistic system. The calculations reveal the influence of various processes on the spin squeezing, such as the AC Stark effect and the spontaneous emission, but also show great agreement with the previous experiments by Z. Chen et al.~\cite{ZChen} and J. G. Bohnet et al.~\cite{JGBohnet}. In the future, equipped with the ability of the modified QuantumCumulant.jl package to automatically derive and solve the stochastic equations, the proposed approach can be extended to address more exotic quantum-measurement effects of large quantum systems, such as deterministic spin squeezing with quantum feedback~\cite{LKThomsen,LKThomsenPRA,KCCox}, spin squeezing of optical clock transitions~\cite{EPP2020,BBraverman}, retrodictive spin squeezing with past quantum states~\cite{HBao,HBaoNC,JMR2024}, enhanced continuous phase estimation with spin squeezing~\cite{AShankar} and so on.

\begin{acknowledgments}
ZhiQing Zhang carried out the numerical calculations under the supervision of Yuan Zhang who developed the theory and the numerical programs. They contribute equally to the work. All authors contributed to the analyses and the writing of the manuscript. This work is supported by the Scientific Research Innovation Capability Support Project for Young Faculty grant SRICSPYF-BS2025008, the National Key R\&D Program of China grant 2024YFE0105200, the National Natural Science Foundation of China under grants 12422413, 12174347, 12074232, 12125406, 62027816, U21A2070, the Beijing National Laboratory for Condensed Matter Physics under the grant 2023BNLCMPKF017, and the Cross-disciplinary Innovative Research Group Project of Henan Province No. 232300421004, as well as by the Carlsberg Foundation through the “Semper Ardens” Research Project QCooL.
\end{acknowledgments}

\appendix
\renewcommand\thefigure{A\arabic{figure}}
\renewcommand\thetable{A\arabic{table}}
\section{Solutions of Standard Stochastic Master Equation \label{sec:standSME}}
In the main text, we have benchmarked the proposed stochastic mean-field approach with an exact stochastic collective density matrix approach by solving the standard stochastic master equation [SME, see Eq.~\eqref{eq:standard-sme} in the main text] for an ensemble of identical two-level atoms subject to a homodyne detection with a measurement strength $M$. In this appendix, we present the equations derived from  the standard SME based on the two approaches, along with the expressions to compute the collective spin vector components and the spin squeezing parameter.

\subsection{Stochastic Mean-field Approach\label{subsec:mean}}
In our proposed stochastic mean-field approach, we solve the standard SME (1) of the main text by deriving equations $\partial_t\langle \hat{o} \rangle=\text{tr}\{(\partial_t \hat{\rho}_I)\hat{o}\}$ for the mean value $\langle\hat{o}\rangle$ of the operator $\hat{o}$, and applying the second-order cumulant expansion approximation $\left \langle\hat{o}\hat{p}\hat{q} \right \rangle = \left \langle \hat{o}\right \rangle \left \langle \hat{p}\hat{q} \right \rangle + \left \langle \hat{p}\right \rangle \left \langle \hat{o}\hat{q} \right \rangle + \left \langle \hat{q}\right \rangle \left \langle \hat{o}\hat{p} \right \rangle - 2 \left \langle \hat{o}\right \rangle \left \langle \hat{p} \right \rangle \left \langle \hat{q} \right \rangle$ for single-particle
and single-photon operators $\hat{o}, \hat{p}, \hat{q}$
to truncate the hierarchy of equations. To be consistent with the convention of the QuantumCumulants.jl package, we label the upper and lower level of the pesudo-spin with the integer 1, 2, and then introduce the projection operator $\hat{\sigma}_k^{22}$, the transition operators $\hat{\sigma}_k^{12}$, $\hat{\sigma}_k^{21}$ of the $k$-th atom. Then, we can write the collective spin vector component operators as $\hat{J}_x = (1/2)\sum_{k=1}^N (\hat{\sigma}_k^{12} + \hat{\sigma}_k^{21}), \hat{J}_y = (i/2) \sum_{k=1}^N(\hat{\sigma}_k^{12} - \hat{\sigma}_k^{21}), \hat{J}_z = (1/2)\sum_{k=1}^N (2\hat{\sigma}_k^{22}-\hat{1}_k)$, and reformulate the standard SME as $\partial_t \hat{\rho}_I = -M\mathcal{D}[\sum_{k=1}^N\hat{\sigma}_k^{22}]\hat{\rho}_I+dW/dt\sqrt{M}\mathcal{H}[\sum_{k=1}^N\hat{\sigma}_k^{22}]\hat{\rho}_I$. 

Following the protocol detailed above, we derive the equations of motion for the expectation values of single-atom operators, including the population $\langle \hat{\sigma}_k^{22}\rangle$ and the coherences $\langle \hat{\sigma}_k^{12}\rangle$ and $\langle \hat{\sigma}_k^{21}\rangle$, as well as for the two-atom correlations $\langle\hat{\sigma}_k^{ij} \hat{\sigma}_{k'}^{lm}\rangle$. Assuming that all the atoms are identical, these single-atom and two-atom expectation values are identical for any atom and any atom pair, respectively. Consequently, the dynamics can be fully characterized by the terms of the first atom and the first pair of atoms. We therefore obtain the following equations. The atomic coherence $\left \langle \hat{\sigma}_1^{12}\right \rangle$ and population $\left \langle \hat{\sigma}_1^{22}\right \rangle$ on the upper state follow the equations

\begin{align}
&\partial_{t}\langle\hat{\sigma}_1^{22}\rangle=\frac{dW}{dt}\sqrt{M}\left[ 2\langle\hat{\sigma}_1^{22}\rangle \right.\nonumber \\
&\left.+2(N-1)\langle\hat{\sigma}_1^{22}\hat{\sigma}_1^{22}\rangle-2N\langle\hat{\sigma}_1^{22}\rangle^2\right],
\label{A1}
\end{align}

\begin{align}
&\partial_{t}\langle\hat{\sigma}_1^{12}\rangle=-M/2\langle\hat{\sigma}_1^{12}\rangle+\frac{dW}{dt}\sqrt{M}\left[ \langle\hat{\sigma}_1^{12}\rangle \right.\nonumber\\
&\left.+2(N-1)\langle\hat{\sigma}_1^{12}\hat{\sigma}_1^{22}\rangle-2N\langle\hat{\sigma}_1^{12}\rangle\langle\hat{\sigma}_1^{22}\rangle\right].
\end{align}
The equations for the atom-atom correlations read
\begin{align}
&\partial_{t}\langle\hat{\sigma}_1^{12}\hat{\sigma}_1^{12}\rangle=-2M\langle\hat{\sigma}_1^{12}\hat{\sigma}_1^{12}\rangle+\frac{dW}{dt}\sqrt{M}\left[ 2\langle\hat{\sigma}_1^{12}\hat{\sigma}_1^{12}\rangle \right.\nonumber\\
& -4\langle\hat{\sigma}_1^{12}\hat{\sigma}_1^{12}\rangle\langle\hat{\sigma}_1^{12}\rangle \nonumber\\
&+4(N-2)(\langle\hat{\sigma}_1^{12}\hat{\sigma}_1^{22}\rangle\langle\hat{\sigma}_1^{12}\rangle-\langle\hat{\sigma}_1^{12}\rangle^2\langle\hat{\sigma}_1^{22}\rangle)],
\end{align}

\begin{align}
&\partial_{t}\langle\hat{\sigma}_1^{12}\hat{\sigma}_1^{21}\rangle=\frac{dW}{dt}\sqrt{M}\left[ 2\langle\hat{\sigma}_1^{12}\hat{\sigma}_1^{21}\rangle-4\langle\hat{\sigma}_1^{12}\hat{\sigma}_1^{21}\rangle\langle\hat{\sigma}_1^{22}\rangle \right.\nonumber\\
& +4(N-2)(\langle\hat{\sigma}_1^{12}\hat{\sigma}_1^{22}\rangle\langle\hat{\sigma}_1^{12}\rangle-\langle\hat{\sigma}_1^{12}\rangle^2\langle\hat{\sigma}_1^{22}\rangle)],
\end{align}

\begin{align}
&\partial_{t}\langle\hat{\sigma}_1^{22}\hat{\sigma}_1^{22}\rangle=\frac{dW}{dt}\sqrt{M}\left[ 4\langle\hat{\sigma}_1^{22}\hat{\sigma}_1^{22}\rangle\right.\nonumber\\
&+4(N-3)\langle\hat{\sigma}_1^{22}\hat{\sigma}_1^{22}\rangle\langle\hat{\sigma}_1^{22}\rangle-4(N-2)\langle\hat{\sigma}_1^{22}\rangle^3],
\end{align}

\begin{align}
&\partial_{t}\langle\hat{\sigma}_1^{12}\hat{\sigma}_1^{22}\rangle=-M/2\langle\hat{\sigma}_1^{12}\hat{\sigma}_1^{22}\rangle+\frac{dW}{dt}\sqrt{M}\left[ 3\langle\hat{\sigma}_1^{12}\hat{\sigma}_1^{22}\rangle\right.\nonumber\\
&\left.+2(N-2)\langle\hat{\sigma}_1^{22}\hat{\sigma}_1^{22}\rangle\langle\hat{\sigma}_1^{12}\rangle\right.\nonumber\\
&+2(N-4)\langle\hat{\sigma}_1^{12}\hat{\sigma}_1^{22}\rangle\langle\hat{\sigma}_1^{22}\rangle-4(N-2)\langle\hat{\sigma}_1^{22}\rangle^2\langle\hat{\sigma}_1^{12}\rangle],
\end{align}

\begin{align}
&\partial_{t}\langle\hat{\sigma}_1^{21}\hat{\sigma}_1^{22}\rangle=-M/2\langle\hat{\sigma}_1^{21}\hat{\sigma}_1^{22}\rangle+\frac{dW}{dt}\sqrt{M}\left[ 3\langle\hat{\sigma}_1^{21}\hat{\sigma}_1^{22}\rangle\right.\nonumber\\
&\left.+2(N-2)\langle\hat{\sigma}_1^{22}\hat{\sigma}_1^{22}\rangle\langle\hat{\sigma}_1^{21}\rangle\right.\nonumber\\
&+2(N-4)\langle\hat{\sigma}_1^{21}\hat{\sigma}_1^{22}\rangle\langle\hat{\sigma}_1^{22}\rangle-4(N-2)\langle\hat{\sigma}_1^{22}\rangle^2\langle\hat{\sigma}_1^{21}\rangle].
\label{A7}
\end{align}
Since other certain quantities, such as $\langle\hat{\sigma}_1^{21}\rangle$, $\langle\hat{\sigma}_1^{21}\hat{\sigma}_1^{21}\rangle$, are complex conjugates of the quantities considered, i.e. $\langle\hat{\sigma}_1^{21}\rangle=\langle\hat{\sigma}_1^{12}\rangle^{\ast}$, $\langle\hat{\sigma}_1^{21}\hat{\sigma}_1^{21}\rangle=\langle\hat{\sigma}_1^{12}\hat{\sigma}_1^{12}\rangle^{\ast}$, we do not consider their equations.

By utilizing the first-order and second-order
mean-fields mentioned above, the mean-fields of the collective spin vector components operators, and the square of these operators can be computed with 
\begin{align}
&J_x=\frac{N}{2}(\langle\hat{\sigma}_1^{12}\rangle+\langle\hat{\sigma}_1^{21}\rangle), \\
&J_y=\frac{iN}{2}(\langle\hat{\sigma}_1^{12}\rangle-\langle\hat{\sigma}_1^{21}\rangle), \\
&J_z=\frac{N}{2}(2\langle\hat{\sigma}_1^{22}\rangle-1),
\end{align}
\begin{align}
&\langle \hat{J}_x^2\rangle=\frac{1}{4}[N+(N^2-N)(\langle\hat{\sigma}_1^{12}\hat{\sigma}_2^{12}\rangle \nonumber \\
&+\langle\hat{\sigma}_1^{12}\hat{\sigma}_2^{21}\rangle+\langle\hat{\sigma}_1^{21}\hat{\sigma}_2^{12}\rangle+\langle\hat{\sigma}_1^{21}\hat{\sigma}_2^{21}\rangle)],  \\
&\langle \hat{J}_y^2\rangle=\frac{1}{4}[N-(N^2-N)(\langle\hat{\sigma}_1^{12}\hat{\sigma}_2^{12}\rangle \nonumber \\
&-\langle\hat{\sigma}_1^{12}\hat{\sigma}_2^{21}\rangle-\langle\hat{\sigma}_1^{21}\hat{\sigma}_2^{12}\rangle+\langle\hat{\sigma}_1^{21}\hat{\sigma}_2^{21}\rangle)], \\
\langle \hat{J}_z^2\rangle&=\frac{1}{4}[4(N^2-N)(\langle\hat{\sigma}_1^{22}\hat{\sigma}_2^{22}\rangle-\langle\hat{\sigma}_1^{22}\rangle)+N^2].
\end{align} 
From the above quantities, we can compute the uncertainty of the collective spin components $\Delta J_i=\sqrt{\langle\hat{J}_i^2\rangle-J_i^2}$ and the spin squeezing parameters $\xi_i^2=\frac{N(\Delta J_i)^2}{\langle\hat{J}_j^2
\rangle+\langle\hat{J}_k^2
\rangle}$ (with $i, j, k=x, y, z$ and $i \neq j\neq k$).

\begin{figure}
\begin{centering}
\includegraphics[scale=0.45]{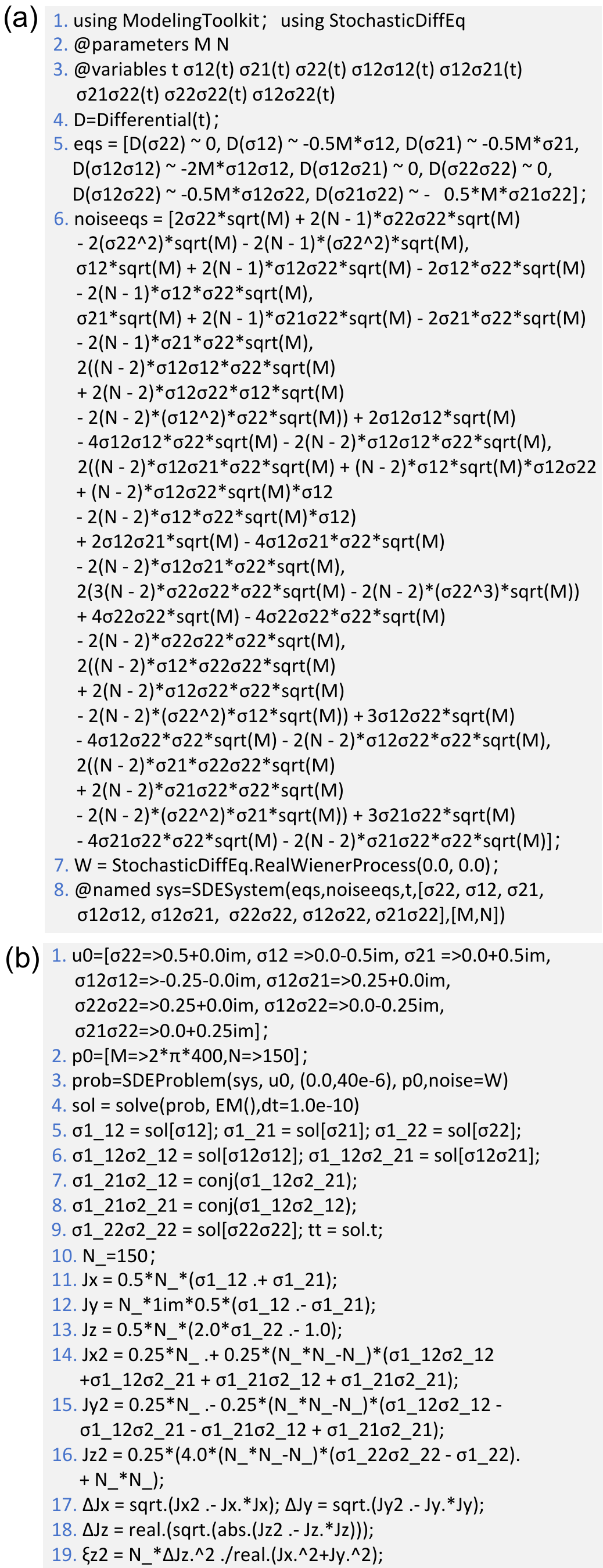}
\par\end{centering}
\caption{Julia codes to construct (a) and solve (b) the stochastic mean-field equations (\ref{A1}-\ref{A7}) derived from the standard SME, and the calculation of the collective spin components and spin squeezing parameter $\xi_z^2$. }
\label{fig:code-stabdard_meanfield}
\end{figure}

Figure.~\ref{fig:code-stabdard_meanfield} shows the Julia codes to solve Eqs.~\eqref{A1} to \eqref{A7}, and to compute the observables of interest. In Fig.~\ref{fig:code-stabdard_meanfield}(a), line 1 imports the packages. Lines 2, 3 declare the parameters of the equations and the time-dependent variables. Lines 4, 5 define the deterministic part of stochastic mean-field equations. Lines 6, 7 define 
the stochastic part of stochastic mean-field equations and the wiener noise. Line 8 defines a stochastic differential equation (SDE) system, including deterministic dynamics, stochastic terms, variables, and parameters for numerical simulation.

In Fig.~\ref{fig:code-stabdard_meanfield}(b), lines 1, 2 define the initial conditions for the atoms prepared in the coherent spin state (CSS), specifying the physical parameters of the system. Line 3 defines the SDE problem with given initial values, time range, parameters, and noise type. Line 4 solves the SDE problem. Line 5 extracts the coherence and population of the two-level system. Lines 6-9 extract the atom-atom correlations. Lines 10-13 calculate the expectation values of the collective spin components. Lines 14-16 calculate the expectation values of the squared collective spin components. Lines 17, 18 calculate the uncertainty of the collective spin components. Line 19 calculates the spin squeezing parameter along the z-direction of the collective spin.

\subsection{Stochastic Collective Density Matrix Approach \label{subsec:coldm}}

The stochastic collective density matrix approach has been detailed in our previous study~\cite{YZhang}. Below, we summarize its essential elements for completeness. To solve the  SME, we can introduce the density matrix elements $\rho_{\alpha\beta}={\rm tr}\{\hat{\rho} \left|\alpha\right\rangle\left\langle\beta\right|\}$ with the product states $\left|\alpha\right\rangle=\bigotimes_{k=1}^{N}\left|l,k\right\rangle$ and $\left|\beta\right\rangle=\bigotimes_{k=1}^{N}\left|l',k\right\rangle$, where the symbols $l,l' =\uparrow, \downarrow$ label the upper and lower levels of the pseudo-spins. Due to the indistinguishability and identical coupling of all pseudo-spins, many elements are equivalent under permutation symmetry. This equivalence allows us to classify the matrix elements uniquely by a set of four occupation numbers $n_{\uparrow\uparrow}, n_{\uparrow\downarrow}, n_{\downarrow\uparrow}, n_{\downarrow\downarrow}$. Here, $n_{l,l'}$ counts how many spins are in state $l$ in $|\alpha\rangle$ and $l'$ state in $|\beta\rangle$, thus satisfies the condition $\sum_{l,l'} n_{l,l'} = N$. Consequently, all symmetry-equivalent elements are mapped to a single collective variable $\langle n\rangle \equiv \Big\langle \begin{array}{cc} n_{\uparrow\uparrow} & n_{\uparrow\downarrow}\\
n_{\downarrow\uparrow} & n_{\downarrow\downarrow}
\end{array}\Big\rangle$, and  the number of independent elements reduces from the exponential scaling $4^N$  to the polynomial scaling $C_N^3 \approx N^3$. 

Following the protocol outlined in Ref~\cite{YZhang},  the SME~\eqref{eq:standard-sme} translates into a closed set of stochastic differential equations for the collective variables:
 \begin{align}
&\partial_t\langle n \rangle=-\frac{M}{2}(n_{\uparrow\downarrow}-n_{\downarrow\uparrow})^2\langle n \rangle \nonumber \\
&+\sqrt{M}\frac{dW}{dt}  \{\left(n_{\uparrow\uparrow}-n_{\downarrow\downarrow}\right) -2\langle \hat{J}_z \rangle\}\langle n \rangle. \label{eq:sme-final}
\end{align}
If we specify the initial state of individual pseudo-spin as $|\psi_k\rangle=c_\uparrow|\uparrow,k\rangle+c_{\downarrow}|\downarrow,k\rangle$ with complex numbers $c_{\uparrow}=\rm{sin}(\theta/2)e^{i\phi}$ and $c_{\downarrow}=\rm{cos}(\theta/2)$, where the angle $\theta$, $\phi$ are the azimuth ad polar angle of the Bloch vector. 
Then, the initial collective density matrix elements can be calculated as 
\begin{align}
\langle n \rangle_0=\prod_{l,l'}(c_{l}c^{\ast}_{l'})^{n_{ll'}}.
\end{align}

After solving the Eq.~(\ref{eq:sme-final}) for the collective density matrix elements, we now calculate the quantities of interest. The mean-fields of the collective operators  can be calculated as 
\begin{align}
&J_{x}=\sum_{l=0}^{N}C_{N}^{l}l {\rm Re}\Big\langle\begin{matrix}l-1 & 0\\ 1 & N-l\end{matrix}\Big\rangle, \\
&J_{y}=-\sum_{l=0}^{N}C_{N}^{l}l {\rm Im}\Big\langle\begin{matrix}{l-1} & {0}\\ {1} & {N-l}\end{matrix}\Big\rangle,  \\
& J_{z}=\frac{1}{2}\sum_{l=0}^{N}C_{N}^{l}\left(2l-N\right)\Big\langle\begin{matrix}{l} & {0}\\ {0} & {N-l}\end{matrix}\Big\rangle.
\end{align}
The mean fields of the squared collective operators can be computed as 
\begin{align}
& \langle\hat{J}_{x}^2\rangle=\frac{1}{4}\sum_{l=0}^{N}C_{N}^{l}\Big[l(l-1)\Big\langle\begin{array}{cc}{l-2} & {0}\\ {2} & {N-l}\end{array}\Big\rangle \nonumber \\
&+N\Big\langle\begin{array}{cc}{l} & {0}\\ {0} & {N-l}\end{array}\Big\rangle+2l(N -l)\Big\langle\begin{matrix}l-1 & 1\\ 1 & N-l-1\end{matrix}\Big\rangle\nonumber \\
&+\left(N-l\right)\left(N-l-1\right)\Big\langle\begin{matrix}l & 2\\ 0 & N-l-2\end{matrix}\Big\rangle\Big], \\
& \langle\hat{J}_{y}^2\rangle=-\frac{1}{4}\sum_{l=0}^{N}C_{N}^{l}\Big[l(l-1)\Big\langle\begin{matrix}{l-2} & {0}\\ {2} & {N-l}\end{matrix}\Big\rangle\nonumber \\
&-N\Big\langle\begin{matrix}{l} & {0}\\ {0} & {N-l}\end{matrix}\Big\rangle-2l(N-l)\Big\langle\begin{matrix}l-1 & 1\\ 1 & N-l-1\end{matrix}\Big\rangle\nonumber\\ 
&+(N-l)(N-l-1)\Big\langle\begin{matrix}l & 2\\ 0 & N-l-2\end{matrix}\Big\rangle\Big],   \\
& \langle\hat{J}_{z}^2\rangle=\frac{1}{4}\sum_{l=0}^{N}C_{N}^{l}(2l-N)^2\Big\langle\begin{matrix}{l} & {0}\\ {0} & {N-l}\end{matrix}\Big\rangle.
\end{align}
From the above quantities, we can calculate the uncertainty of the collective spin components with $\Delta J_i =\sqrt{\langle \hat{J}_i^2\rangle - \langle \hat{J}_i\rangle^2}$ and the spin squeezing parameters $\xi_i^2=\frac{N(\Delta J_i)^2}{\langle\hat{J}_j^2
\rangle+\langle\hat{J}_k^2
\rangle}$ (with $i, j, k=x, y, z$ and $i \neq j\neq k$).

We have written a program based on CUDA language, which allows us to use the NIVIDA graphic card to carry out the parallel computation. With such a program, we estimate that the simulations of 100 pseudo-spins can be finished in few minutes, and the simulation with up to 1000 pseudo spins is possible. We have uploaded the program in the GitLab~\cite{code}, and the results shown in the main text can be obtained by modifying the simulation parameters in the file “parameters.cu“.

\section{\label{sec:code} Julia Codes to Automatically Derive and Solve Stochastic Mean-field Equations}

\begin{figure}
\begin{centering}
\includegraphics[scale=0.43]{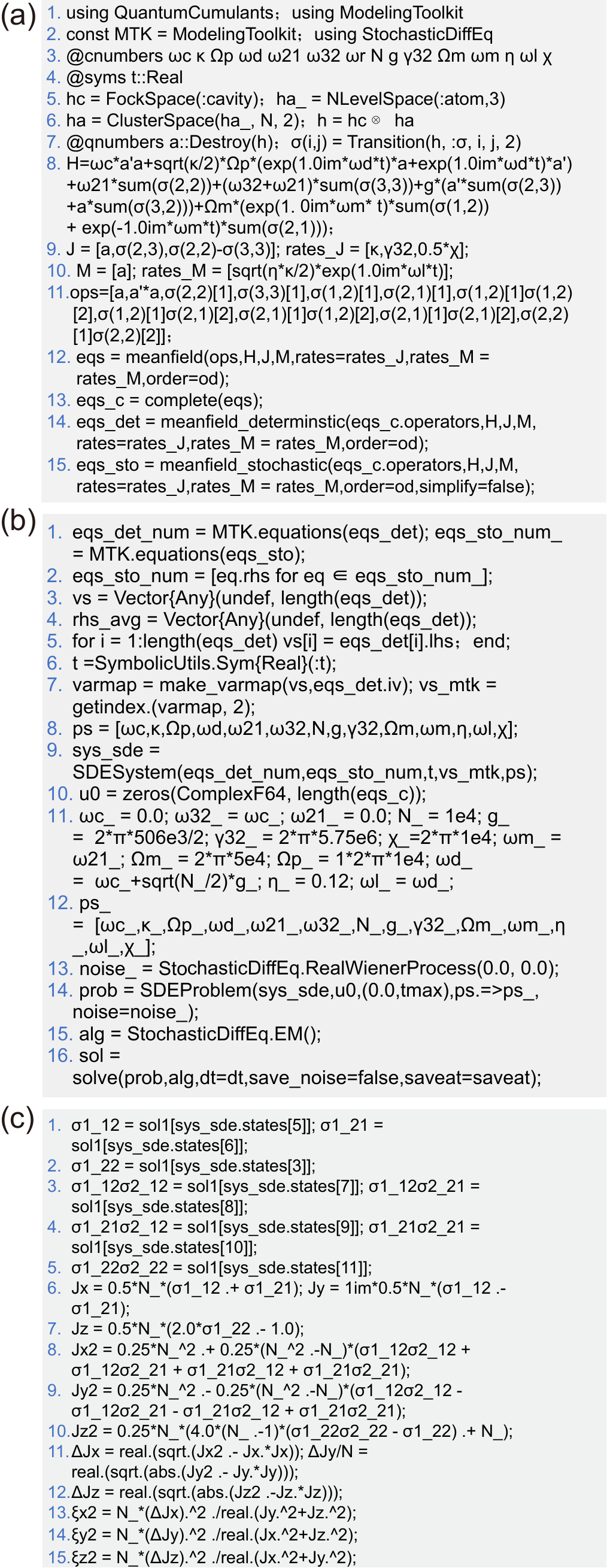}
\par\end{centering}
\caption{Julia codes to derive (a), solve (b) the stoachastic mean-field equations, and to extract the numerical results (c).}
\label{fig:julia-sto}
\end{figure}

\begin{figure}
\begin{centering}
\includegraphics[scale=0.46]{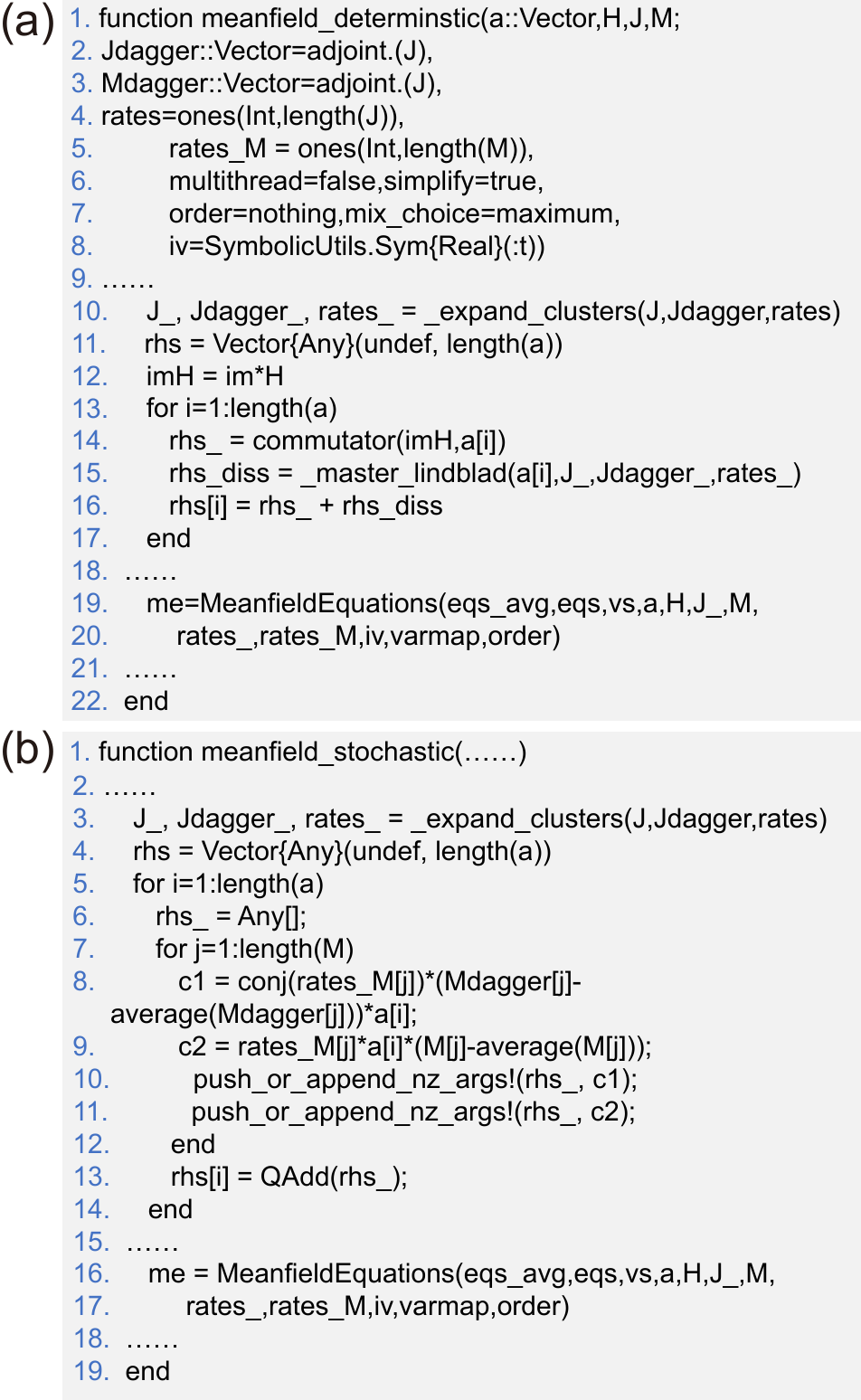}
\par\end{centering}
\caption{Partial codes of the “meanfield\_deterministic” function (a) and the “meanfield\_stochastic” function (b). } 
\label{fig:mf_det_sto}
\end{figure}

In this Appendix, we introduce the Julia codes used to derive (a) and solve (b, c) the stochastic mean-field equations of realistic system in Fig.~\ref{fig:julia-sto}, as well as the partial modified codes for the QuantumCumulants.jl package shown in Fig.~\ref{fig:mf_det_sto}.  It is worth of noticing that the same functionality will be incorporated in the next version of that package, and it would be more convenient to reproduce the results presented in the current study. 

In Fig.~\ref{fig:julia-sto}(a),  lines 1, 2 import the packages. Lines 3, 4 define complex quantities and time argument as real number. Line 5 defines the Hilbert space of the optical cavity, and a three-level atom. Line 6 defines the Hilbert space for the atomic ensemble, and the product Hilbert space for the atoms-cavity system. Line 7 defines the photon annihilation operator and the atomic transition operators. Line 8 defines the system Hamiltonian. Lines 9 and 10 define the list of operators and rates for the Lindblad terms and the measurement backaction, respectively. Line 11 defines the list of operators, line 12  derives the equations for the mean-value of these operators, line 13 derives the complete set of mean-field equations. Lines 14 and 15 define the equations for the deterministic and stochastic dynamics. 

In Fig.~\ref{fig:julia-sto}(b), lines 1 and 2 define the numerical version of the equations for the deterministic and stochastic dynamics. Lines 3-7 define the list of variables. Line 8 defines the list of complex numbers. Line 9 defines the SDE system. Line 10 defines the initial values for the mean field. Lines 11 and 12 specify the values for the complex numbers, and the list of these values. Line 13 defines the Wiener noise for the stochastic dynamics. Line 14 defines the SDE problem with given initial values,  time range, parameters, and noise type.  Line 15 specifies the Euler-Maruyama solver, and the list of simulation times. Line 16 solves the SDE problem.

In Fig.~\ref{fig:julia-sto}(c), lines 1, 2 extract the  coherence of the transition between the upper hyper-fine ground state and the excited state, and the population of the upper hyper-fine ground state. Lines 3-5 extract the atom-atom correlations. Lines 6, 7 calculate the expectation values of the collective spin operators. Lines 8-10 calculate the expectation values of the squared collective spin operators. Lines 11, 12  calculate the uncertainty of the collective spin components. The derived equation is shown in Sec.~\ref{sec:meaneq}.

While the QuantumCumulants.jl package is designed to solve the standard and deterministic quantum master equation, we have modified it to solve the SME. In the modification, the most significant change is to define a new function “meanfield\_stochastic” to derive the dependence of the mean-field quantities due to the measurement backaction.

In Fig.~\ref{fig:mf_det_sto}(a), we show the partial code for the “meanfield\_deterministic” as a reference, which resembles the “meanfield” function in the QuantumCumulants.jl package. Line 1 defines the function with the necessary parameters. The first two parameters are the list of operators “a” to define the mean field quantities and the system Hamiltonian “H”. The next two parameters are the list of operators “J, M” to define the Lindblad and stochastic term. Lines 2, 3 define the conjugation “Jdagger, Mdagger” of “J, M”. Lines 4, 5 are the rates “rates, rates\_M” associated with the Lindblad and stochastic term. Lines 6-8 define the remaining parameters. Among them, “order,iv” are the order of the mean-field approach, and the symbol for the time parameter. Line 10 translates the parameters “J, Jdagger” in Hilbert space to deal with the system within the identical particle assumption. Line 11 defines the array of operators with the same length as the list “a”.  Line 12 defines “imH” as the multiplication of the imaginary sign with the system Hamiltonian for convenience. Lines 13-17 compute the dependence due to the commutation relation of “a” and “imH”, and due to the Lindblad term, and form the equations for the operators in the list “a”. We transfer these equations into  “eqs\_avg” for the mean-field quantities by using the “average()” function (not shown). Lines 19-20 define the object “me” to represent the derived mean-field equations. One should note that in this function the lists “M,Mdagger, M\_rates” for the stochastic terms are used only for the definition of the object “me”. 

In Fig.~\ref{fig:mf_det_sto}(b), we show the partial code for the “meanfield\_stochastic” function. In comparison to the “meanfield\_deterministic” function, only the code lines 5-14 are different. Here, we define first an empty list “rhs\_”, and we then enumerate every term in the list “M”, and compute the dependence “c1, c2” due to the corresponding stochastic terms, and push them into the list “rhs\_”, and finally add the terms in the list to form the equations for the operator in the list “a”. Here, the most important step is to introduce the mean-field terms through the “average()” functions into the equations for the operators.  In lines 16-20, we define again the object “me” to represent the derived mean-field equations. One should note that in this function the lists “J,Jdagger,J\_rates” for the stochastic terms are used only for the definition of the object “me”.  

In our code, a single cavity mode is treated as a component of the system. Often, one may eliminate the cavity mode and deal only with an SME for the atoms, where the cavity coupling enters through collective energy shifts and decay rates. We have verified that the above modified codes can also be applied to solve this simplified equation.

\section{Supplemental Results \label{sec:supresult}}
In this Appendix, we provide extra results to complement the discussions in the main text. 

\subsection{System Parameters}

Here, we detail the system parameters for the simulations in the main text and this supplemental material, see also Tab.~\ref{tab:params}.
\begin{table}
\footnotesize
\caption{System parameters for the simulations}\label{tab:params}
\doublerulesep 0.1pt \tabcolsep 15pt 
\begin{tabular}{ccc}
\toprule
  System Parameters & Value\\\hline
  Cavity mode frequency $\omega_{c}$ & $2\pi\times377$ THz \\
  $\omega_{21}$ & $2\pi\times 6.8$ GHz\\
  Photo damping rate $\kappa$ & $2\pi\times11.1$ MHz\\
  $\omega_{32}$ & $\omega_{c}$\\
  Atom-cavity couple strength $g$ & $2\pi \times 0.253 $ MHz\\
  Probe field frequency $\omega_{p}$ & $\omega_{c} + \sqrt{N/2}g$\\
  Decay rate $\gamma$ & $2\pi\times5.75$ MHz\\
  Probe field strength $\Omega_p$ & $2\pi\times 10^4$ $\rm \sqrt{\rm Hz}$\\
  microwave field strength $\Omega_m$ & $2\pi\times 10^6$ $\rm \sqrt{\rm Hz}$\\
  microwave field frequency $\Omega_p$ & $\omega_{21}$\\
  Atom number $N$ & $10^{4}$\\
  Detection effciency $\eta$ & $0.12$\\
  Dephasing rate $\chi$ & $2\pi \times10$ kHz\\
\bottomrule
\end{tabular}

\end{table}
The  optical cavity mode has a frequency $\omega_c=2\pi \times 377 $ THz, and a photon damping rate $\kappa = 2\pi \times 11.1$ MHz. The atomic transition between the upper hyper-fine ground state and the excited state is resonant with the cavity, with a coupling strength $g=2\pi \times 0.253$ MHz, a decay rate $\gamma=2\pi\times 5.75$ MHz and a dephasing rate $\chi =2\pi\times 10$ kHz. The atomic transition between the two hyper-fine ground states has a frequency $\omega_{21} = 2\pi \times 6.8$ GHz, and couples resonantly with the microwave drive field, i.e. $\omega_m =\omega_{21}$, with a strength $\Omega_m = 2\pi\times 10^6$ $\rm \sqrt{\rm Hz}$. In addition, the optical cavity is probed by a laser with a frequency $\omega_p = \omega_c + \sqrt{N/2}g$ and a strength $\Omega_p = 2\pi \times 10^4$ $\rm \sqrt{\rm Hz}$, and the photodetector of the homodyne detection has a detection efficiency $\eta = 0.12$.

\begin{figure}
\begin{centering}
\includegraphics[scale=0.26]{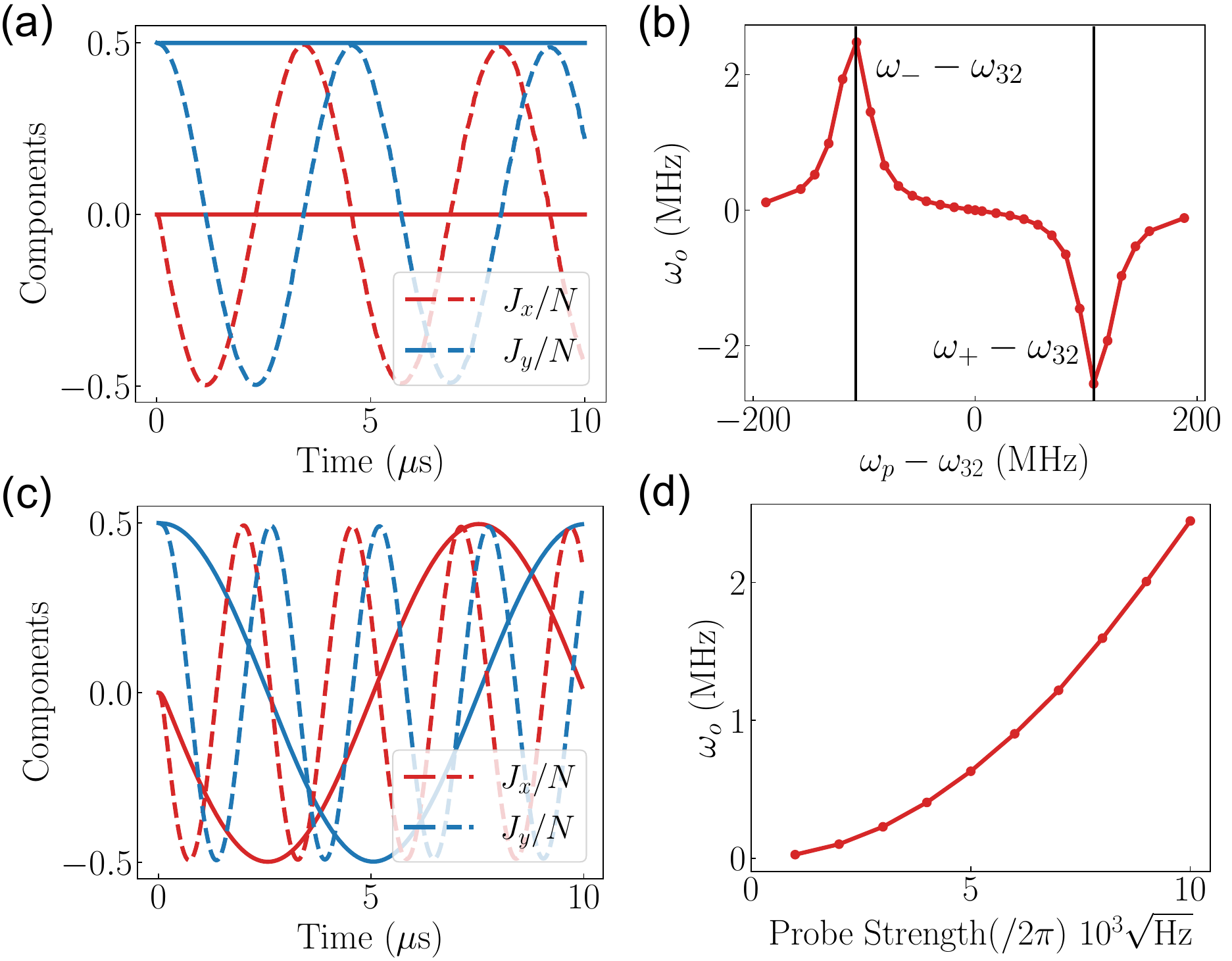}
\par\end{centering}
\caption{ Collective spin vector rotation due to AC Stark effect. (a) shows the dynamics of the collective spin components $J_x$ (red lines) and $J_y$ (blue lines), which are normalized to the number of atoms $N$, for the probe field with frequency detuned ($\omega_p=\omega_{32}+2\pi\times 20$ MHz, dashed lines) and resonant ($\omega_p=\omega_{32}$, solid lines) with respect to the atomic transition between the upper hyper-fine ground state and the excited state, and given field strength $2\pi\times 10^4 \sqrt{\rm Hz}$. (b) shows the oscillation frequencies $\omega_o$ of $J_x$ and $J_y$ as function of the probe field frequency (relative to the the atomic transition frequency $\omega_{32}$) for given probe field strength, which shows resonances at the frequencies $\omega_\pm$ of the atom-photon dressed states. (c) shows the dynamics of $J_x$ and $J_y$ for a strong probe field with strength $2\pi\times 10^4 \sqrt{\rm Hz}$ (dashed lines) and a relatively weak one $2\pi\times 0.5\times10^4 \sqrt{\rm Hz}$ (solid lines), which is resonant with the atoms-photon dressed state with larger frequency, i.e. $\omega_p=\omega_+$. (d) shows the oscillation frequencies $\omega_o$ of $J_x$ and $J_y$ as function of the probe field strength, which can be well fitted with a quadratic function. In all the simulations, the atomic transition is assumed to be resonant with the cavity mode, i.e. $\omega_{32}=\omega_c$. }
\label{fig:rotation}
\end{figure}

\subsection{Collective Spin Vector Rotation due to AC Stark-Shift}

In the main text, we show that the collective spin vector rotates in the equatorial plane during the laser probing of the atomic ensemble. To elucidate this mechanism in detail, we analyze the dependence of the collective spin dynamics on the probe detuning and strength in Fig.~\ref{fig:rotation}. We show in Fig.~\ref{fig:rotation} that this rotation depends on the frequency and strength of the probe field. The rotation occurs only for a probe field detuned from the atomic transition between the upper hyper-fine ground state and the excited state [Fig.~\ref{fig:rotation}(a)], and the frequency of oscillations shows two maxima when the probe frequency is resonant with the two atoms-photon dressed states [Fig.~\ref{fig:rotation}(b)]. In addition, the oscillations become faster when the probe field strength is increased [Fig.~\ref{fig:rotation}(c)], and the oscillation frequency scales quadratically with the probe field strength [Fig.~\ref{fig:rotation}(d)]. The oscillations are readily explained by the AC Stark frequency shift of the ground state of the coupled atoms-cavity system by $\delta_{ac}=-\Omega_p^2/\Delta_1 - \Omega_p^2/\Delta_2$, where $\Delta_1,\Delta_2$ are the frequency detunning of the probe field and the lower and upper dressed state, respectively.

\subsection{Influence of Detection Efficiency and Atoms-cavity Frequency Detuning on Conditional Spin Squeezing~\label{sec:InfDetEff}}

\begin{figure}
\begin{centering}
\includegraphics[scale=0.28]{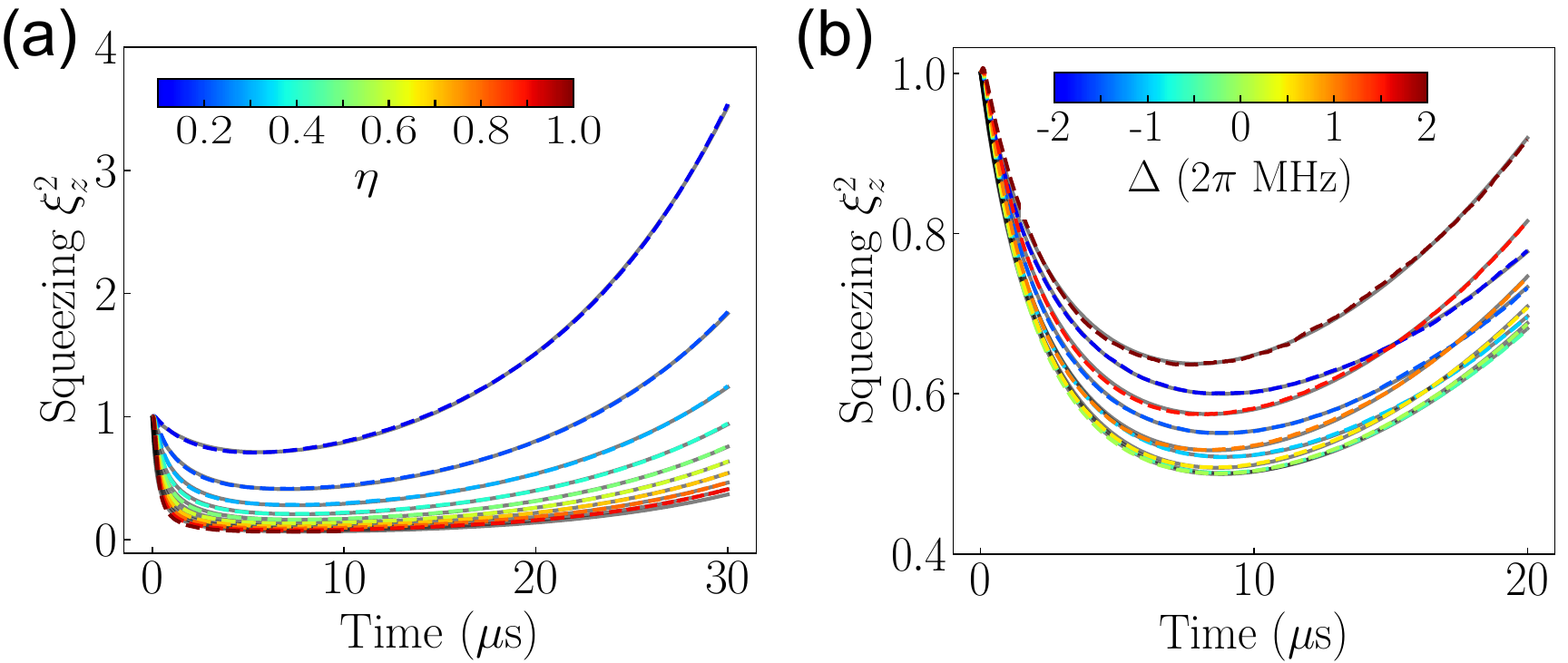}
\par\end{centering}
\caption{ Influence of detection efficiency $\eta$ of photo-detectors (a) and the atoms-cavity frequency detuning $\Delta = \omega_{32}-\omega_c$ (b) on the dynamics of conditional spin squeezing. In these simulations, we consider the fixed probe strength $\Omega_{p}=2\pi\times10^{4} \rm \sqrt{\rm Hz}$ and the presence of the atomic decay with rate $\gamma=2\pi\times5.75$ $\rm MHz$. Other parameters are same as those in Fig. 3(a) of the main text. }
\label{fig:detection}
\end{figure}

In the simulations of the main text, we have assumed a realistic detector efficiency $\eta =0.12$ and a resonant condition $\Delta = \omega_{32} - \omega_c=0$ between the upper hyperfine ground-excited state transition and the cavity mode. In Fig.~\ref{fig:detection}, we investigate how the efficiency $\eta$ (a) and the frequency detuning $\Delta$ (b) affect the conditional spin squeezing. As $\eta$ increases from $0.1$ to $1.0$, the reduction of the initial squeezing parameter becomes much faster, and its minimum becomes gradually lower, while its final increase becomes more slow. As $\Delta$ increases from negative to positive value within several megahertz, the initial reduction of the squeezing parameter becomes first faster and then slower, and a similar behavior is observed for the anti-squeezing at later times. 
\noindent

\begin{figure}
\begin{centering}
\includegraphics[scale=0.28]{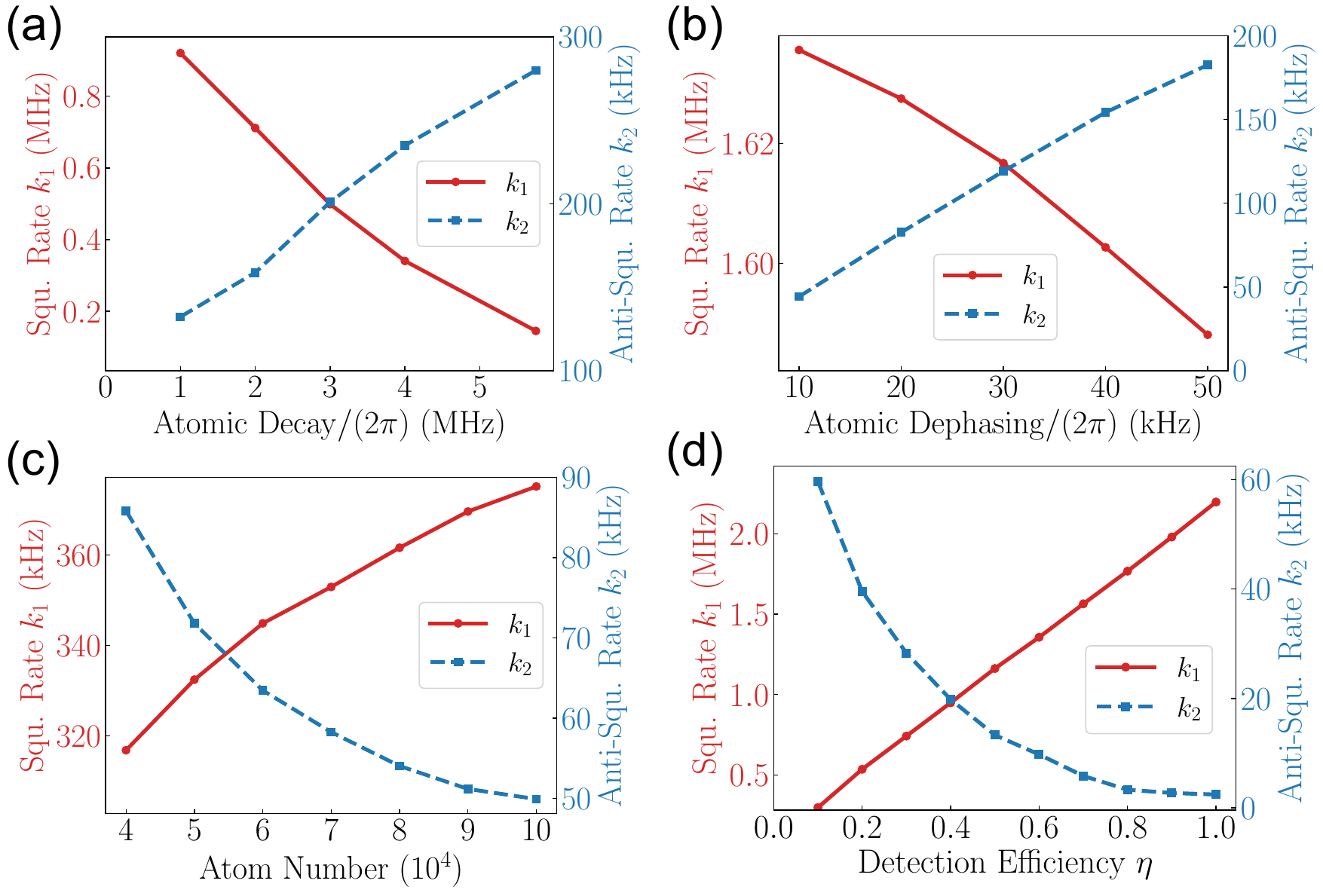}
\par\end{centering}
\caption{ Influence of different parameters on squeezing rate $k_1$ (left axes) and anti-squeezing rate $k_2$ (right axes). (a) and (b) show the results for increasing atomic decay rate $\gamma$, and atomic dephasing rate $\chi$, respectively. (c) and (d) show the results for increasing atom number $N$, and increasing detection efficiency $\eta$, respectively. Other parameters are the same as Tab.~\ref{tab:params}.}
\label{fig:verif}
\end{figure}

\subsection{ Optimal Squeezing and Squeezing Rate}

We study the influence of the atomic decay, the atomic dephasing, the number of atoms on spin squeezing, and discuss the effect of the detection efficiency of the photo-detectors in Appendix.~\ref{sec:InfDetEff}. In the presence of the individual atomic dissipation, the spin squeezing parameter first decreases and then increases in time, which can be well fitted by $\xi_z^2(t)=A/(1+k_1 t) + (1-A)e^{k_2 t}$. From this expression, we extract the squeezing rate $k_1$, the anti-squeezing rate $k_2$, the minimal squeezing parameter $\xi^2_z(\tau)$ and the corresponding optimal time $\tau$, where $\tau=2W(s)/k_2-1/k_1$ with $s\equiv[k_2/(2k_1)]\sqrt{\frac{A}{1-A}\frac{k_1}{k_2}e^{\frac{k_2}{k_1}}}$ and $W(s)$ being the Lambert W function. 

Figure.~\ref{fig:verif} shows the dependence of $k_1$ and $k_2$ on different system parameters. As the decay rate $\gamma$ and the dephasing rate $\chi$ of the individual atom increase, the squeezing rate $k_1$ decreases while the anti-squeezing rate $k_2$ increases [Fig.~\ref{fig:verif}(a,b)]. In contrast, increasing atom number $N$ and detection efficiency of photo-detectors $\eta$ enhances the squeezing rate $k_1$ and suppresses the anti-squeezing rate $k_2$ [Fig.~\ref{fig:verif}(c,d)], demonstrating that the detrimental effects of dissipation can be partially compensated by increasing the atom number and improving the detection efficiency.

\begin{figure}
\begin{centering}
\includegraphics[scale=0.24]{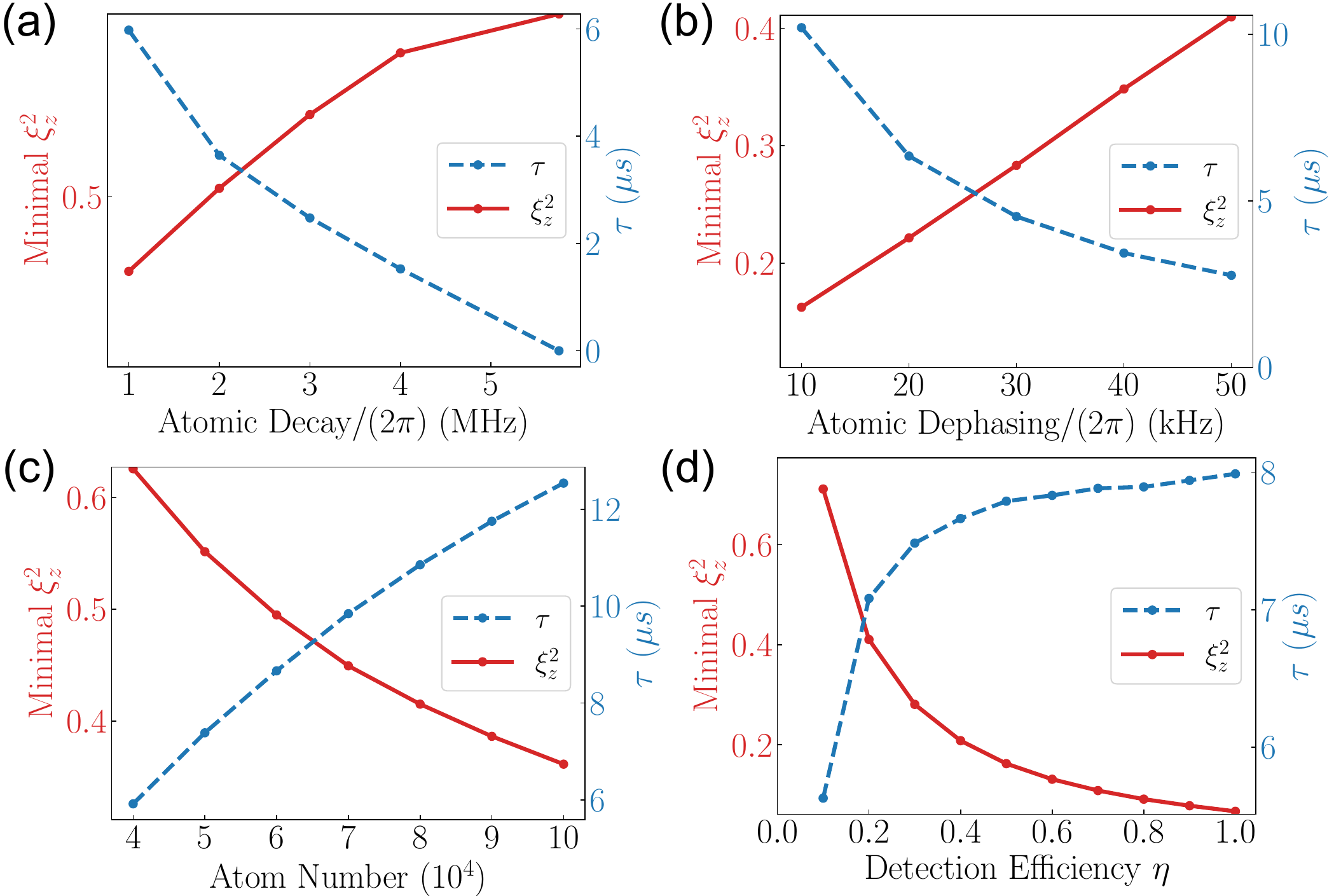}
\par\end{centering}
\caption{ Influence of different parameters on minimal squeezing parameter $\xi_z^2(\tau)$ (left axes) and the corresponding time $\tau$ (right axes). (a) and (b) show the results for increasing atomic decay rate $\gamma$, and  atomic  dephasing rate $\chi$, respectively. (c) and (d) show the results for increasing atom number $N$, and increasing detection efficiency $\eta$, respectively. Other parameters are the same as Tab.~\ref{tab:params}.} 
\label{fig:minsqu}
\end{figure}

Figure.~\ref{fig:minsqu} further summarizes the dependence of the minimal squeezing parameter $\xi^2_z(\tau)$. As the individual atomic decay rate $\gamma$ increases within a few MHz,  $\xi^2_z(\tau)$ increases from about $0.3$ to $1.0$, while $\tau$ decreases from $6$ $\mu$s to nearly zero [Fig.~\ref{fig:minsqu}(a)]. As the individual atomic dephasing rate $\chi$ increases within tens of kHz,  $\xi^2_z(\tau)$ increases from about $0.1$ to $0.4$, and $\tau$ decreases from $10$ $\mu$s to $3$ $\mu$s [Fig.~\ref{fig:minsqu}(b)]. Thus, the atomic dephasing deteriorates the spin squeezing more strongly than decay, see also Fig.~\ref{fig:con_spin_squeezing}. In contrast, as the number of atoms $N$ increases from $4\times 10^4$ to $10^5$,  $\xi^2_z(\tau)$ decreases from about $0.6$ to about $0.35$, and $\tau$ increases from $6$ $\mu$s to $12$ $\mu$s  [Fig.~\ref{fig:minsqu}(c)]. As the detection efficiency $\eta$ increases from $0.12$ to unity, $\xi^2_z(\tau)$ decreases exponentially from about $0.8$ to $0.1$, and $\tau$ increases from $5.5$ $\mu$s to the saturated value $8$ $\mu$s  [Fig.~\ref{fig:minsqu}(d)].

\subsection{Simulation of the Experimental Protocol}

\begin{figure}
\begin{centering}
\includegraphics[scale=0.28]{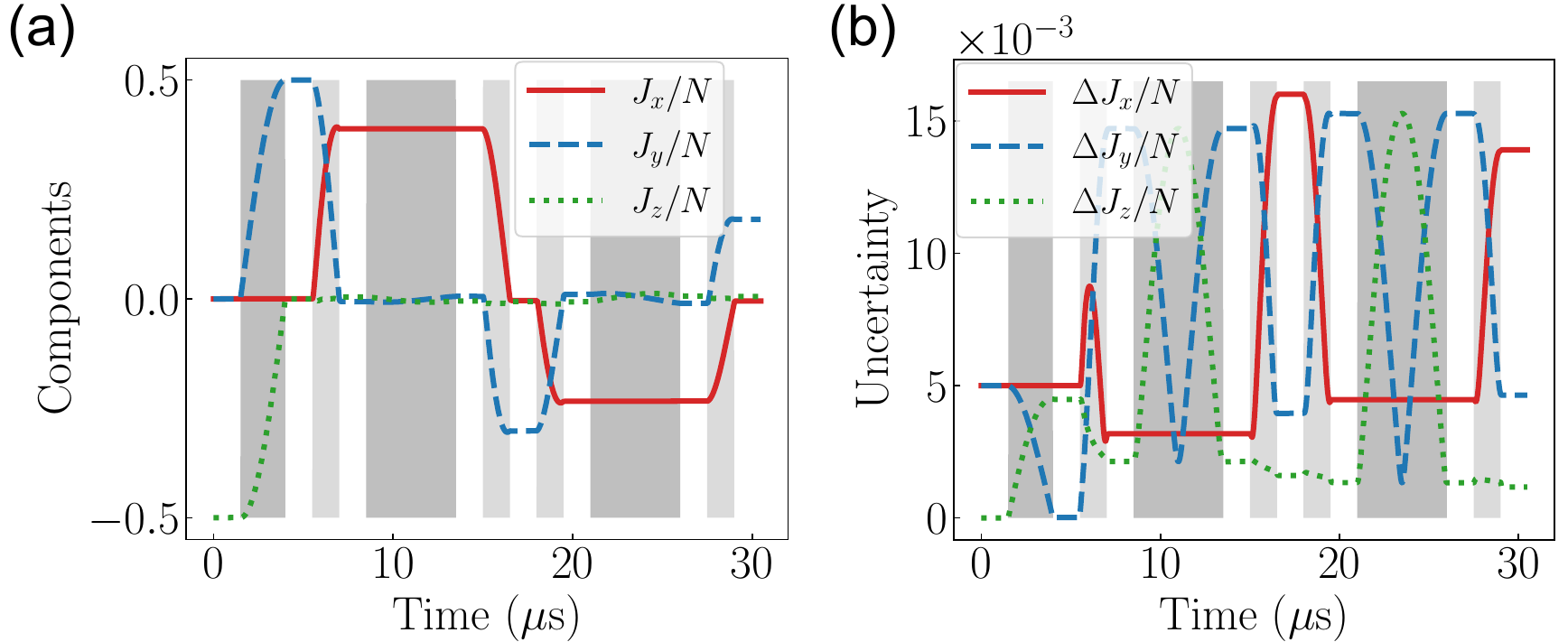}
\par\end{centering}

\caption{ Simulation of the conditional spin squeezing. (a) shows the dynamics of the collective spin components $J_x/N, J_y/N, J_z/N$ (red solid, blue dashed, green dotted line) during the application of three microwave field pulses (gray areas) and four laser probe pulses (light areas). (b) shows the uncertainty of the collective spin components $\Delta J_x/N, \Delta J_y/N, \Delta J_z/N$ (red solid, blue dashed, green dotted line).}
\label{fig:experment}
\end{figure}

In Fig. 4 of the main text, we simulate the protocol implemented in experiment~\cite{ZChen}, demonstrating the evolution of the spin squeezing parameter and photocurrent. In the following, we present additional data on the dynamics of collective spin components and their uncertainties (Fig.~\ref{fig:experment}). We apply first a microwave $\pi/2$-pulse to prepare the atomic ensemble in a CSS, and then we apply two laser probe pulses intersected by a microwave $\pi$-pulse to detect the population of hyper-fine ground states and generate a spin squeezed state, and finally the second step is  repeated to perform measurements on the squeezed state. 

In Fig.~\ref{fig:experment}, we demonstrate the system dynamics during the whole procedure as mentioned above. Fig.~\ref{fig:experment}(a) and (b) show the dynamics of the collective spin components and their uncertainty. We see that after the first microwave $\pi/2$-pulse, the collective spin rotates around the x-axis and becomes parallel with the y-axis, and acquires equal uncertainty along the z-axis and the equatorial plane. During the remaining two microwave $\pi$-pulses, the collective spin is rotated by $180$ degree around the x-axis while its uncertainty is identical before and after the $\pi$-pulses. During the laser probe pulses, the collective spin rotates in the equatorial plane, the uncertainty reduces for the projection along the z-axis, but increases for the projection in the equatorial plane. Note that the projection along the z-axis reduces also slightly during the application of the laser probe pulses.    
 
\begin{table*}
\fontsize{7.6pt}{12pt}\selectfont
\renewcommand{\arraystretch}{2.5}
\setlength{\tabcolsep}{1pt}
\caption{Comparison of different approaches to solve stochastic master equations} \label{tab:approaches}
\begin{tabular}{cccccc}
\toprule
\addlinespace[0ex]
\diagbox{Properties}{Approaches} & \makecell{Standard\\ density matrix} & \makecell{Collective\\ density matrix$^{\textsuperscript{\cite{YZhang}}}$} &  \makecell{Density matrix\\ in Dicke\\ states space$^{\textsuperscript{\cite{MACRossi2020}}}$} & \makecell{\color[RGB]{144, 0, 33}  Stochastic \\ \color[RGB]{144, 0, 33}mean-field } & \makecell{Gaussian-state \\ formulism$^{\textsuperscript{\cite{BLMadsen}}}$} \\[-0.ex]
\midrule 
\addlinespace[0.1ex]
Exact or approximated & Exact & Exact & Exact & \color[RGB]{144, 0, 33}Approximated & Approximated \\[-0.1ex]

Individual atomic dissipation & Yes & Yes & Yes & \color[RGB]{144, 0, 33}Yes & Partially \\[-0.1ex]
Number of levels $l$ & $\geq$2 & $\geq$2 & 2 & \color[RGB]{144, 0, 33}$\geq$2 & 2 \\[-0.1ex]
\makecell{Independent elements \\ (Assuming $l=2$)} & $4^N$ & $N^3$ & $N^3$ & \cellcolor{red!15} \makecell{\color[RGB]{144, 0, 33}  $N^2$ for different atoms;\\
\color[RGB]{144, 0, 33}Independent for \\ \color[RGB]{144, 0, 33}identical atoms} & \makecell{Independent for \\ identical atoms} \\[-0.1ex]
\makecell{Number of simulated  atoms $N$ \\ (Assuming $l=2$)} & \makecell{Few atoms  \\ with \\ different properties} & \makecell{Hundreds of  atoms  \\ with  \\identical properties} & \makecell{Hundreds of  atoms  \\ with  \\identical properties} &  \makecell{\color[RGB]{144, 0, 33} (1)Thousands of atoms \\ \color[RGB]{144, 0, 33}with different properties; \\ \color[RGB]{144, 0, 33} (2)Unlimited for atoms \\\color[RGB]{144, 0, 33}with identical properties} & \makecell{(1)Thousands of atoms \\ with different properties;\\ (2) Unlimited for atoms \\ with identical properties} \\[-0.1ex]
\makecell{Inclusion of atom-atom \\pair coupling} & Possible & Excluded & Excluded & \color[RGB]{144, 0, 33}Possible & Possible \\[-0.1ex]
Parallel computation & - & \makecell{Laptop with\\ NVIDA GPU} & Computer Cluster & \color[RGB]{144, 0, 33}- & - \\[0.0ex]
\bottomrule
\end{tabular}
\end{table*}

\section{Comparison of Different Approaches to Solve Stochastic Master Equation \label{sec:compare}}

In this Appendix, we provide a detailed comparison between our approach and other approaches in Tab.~\ref{tab:approaches} to solve the SME, highlighting the differences and advantages of the proposed approach. Our recently developed approach based on the collective density matrix shares similar properties with the approach based on density matrix in Dicke states space~\cite{YZhang,MACRossi2020}. However, this approach can be potentially applied to the atoms with multiple levels by following the protocol in Ref.~\cite{YZhang}, while the latter approach can only address two-level atoms. 

The stochastic mean-field approach we develop in the current study shares many similarities with the Gaussian state formalism~\cite{BLMadsen}, except that the former can be easily applied to the atoms with multiple levels, and can also account for accurately the individual atomic dissipation. 
In addition, the stochastic mean-field approach outperforms exact solution methods in multiple aspects, such as the consideration of multi-level atoms and a large number of atoms.

\section{Stochastic Mean-field Equations for Realistic System \label{sec:meaneq}}

In this section, we present the mean-field equations derived from the extended SME~\eqref{eq:sme} in the main text with the Julia codes given in Fig.~\ref{fig:julia-sto}. 

We present firstly the equations for the first-order mean quantities. The cavity field amplitude $\langle\hat{a}\rangle$ satisfies the equation
\begin{align}
& \partial_{t}\langle\hat{a}\rangle= -i\left(\omega_{c}-i\kappa/2\right)\langle \hat{a}\rangle+iNg\langle\hat{\sigma}_{1}^{23}\rangle\nonumber \\
& - i\Omega_{p}\sqrt{\kappa/2}e^{-i\omega_{p}t}\nonumber  + \frac{dW}{dt}\sqrt{\eta\kappa/2}e^{i\omega_{p}t}\left(\langle \hat{a}\hat{a}\rangle-\langle \hat{a}\rangle^{2}\right)\nonumber \\
& + \frac{dW}{dt}\sqrt{\eta\kappa/2}e^{-i\omega_{p}t}\left(\langle \hat{a}^{\dagger}\hat{a}\rangle-\langle\hat{a}^{\dagger}\rangle\langle \hat{a}\rangle\right).
\end{align}
The atomic coherence $\left \langle \hat{\sigma}_1^{12}\right \rangle$, $\left \langle \hat{\sigma}_1^{13}\right \rangle$ and $\left \langle \hat{\sigma}_1^{23}\right \rangle$ follow the equations
\begin{align}
& \partial_{t}\langle\hat{\sigma}_1^{12}\rangle=(-i\omega_{21}-\chi/4)\langle\hat{\sigma}_1^{12}\rangle-ig\langle\hat{a}^{\dagger}\hat{\sigma}_1^{13}\rangle\rangle\nonumber \\
& + i\Omega_me^{-i\omega_mt}\left(-1+\langle\hat{\sigma}_1^{33}\rangle+2\langle\hat{\sigma}_1^{22}\rangle\right)\nonumber \\
& + \frac{dW}{dt}\sqrt{\eta\kappa/2}e^{i\omega_{p}t}\left(\langle\hat{a}\hat{\sigma}_1^{12}\rangle - \langle\hat{a}\rangle\langle\hat{\sigma}_1^{12}\rangle\right)\nonumber \\
& + \frac{dW}{dt}\sqrt{\eta\kappa/2}e^{-i\omega_{p}t}\left(\langle\hat{a}^{\dagger}\hat{\sigma}_1^{12}\rangle - \langle\hat{a}^{\dagger}\rangle\langle\hat{\sigma}_1^{12}\rangle\right),
\end{align}
\begin{align}
& \partial_{t}\langle\hat{\sigma}_1^{13}\rangle=\left(-i\omega_{21}-i\omega_{32}-\gamma/2-\chi/4\right)\langle\hat{\sigma}_1^{13}\rangle\nonumber \\
&  - ig\langle\hat{a}\hat{\sigma}_1^{12}\rangle+i\Omega_me^{-i\omega_mt}\langle\hat{\sigma}_1^{23}\rangle\nonumber \\
& + \frac{dW}{dt}\sqrt{\eta\kappa/2}e^{i\omega_{p}t}\left(\langle\hat{a}\hat{\sigma}_1^{13}\rangle - \langle\hat{a}\rangle\langle\hat{\sigma}_1^{13}\rangle\right)\nonumber \\
& + \frac{dW}{dt}\sqrt{\eta\kappa/2}e^{-i\omega_{p}t}\left(\langle\hat{a}^{\dagger}\hat{\sigma}_1^{13}\rangle - \langle\hat{a}^{\dagger}\rangle\langle\hat{\sigma}_1^{13}\rangle\right),   
\end{align}
\begin{align}
& \partial_{t}\langle\hat{\sigma}_1^{23}\rangle=\left(-i\omega_{32}-\gamma/2-\chi\right)\langle\hat{\sigma}_1^{23}\rangle \nonumber \\
& + ig\left(\langle\hat{a}\hat{\sigma}_1^{33}\rangle-\langle\hat{a}\hat{\sigma}_1^{22}\rangle\right)+i\Omega_me^{i\omega_mt}\langle\hat{\sigma}_1^{13}\rangle\nonumber \\
& + \frac{dW}{dt}\sqrt{\eta\kappa/2}e^{i\omega_{p}t}\left (\langle\hat{a}\hat{\sigma}_1^{22}\rangle - \langle\hat{a}\rangle\langle\hat{\sigma}_1^{22}\rangle\right)\nonumber \\
& + \frac{dW}{dt}\sqrt{\eta\kappa/2}e^{-i\omega_{p}t}\left(\langle\hat{a}^{\dagger}\hat{\sigma}_1^{23}\rangle - \langle\hat{a}^{\dagger}\rangle\langle\hat{\sigma}_1^{23}\rangle\right). 
\end{align}
Here and in the following, we do not consider the quantities like $\left \langle \hat{a}^\dagger \right \rangle, \left \langle \hat{\sigma}_1^{21}\right \rangle$ because they are the complex conjugation of other terms like $\left \langle \hat{a} \right \rangle, \left \langle \hat{\sigma}_1^{12}\right \rangle$. The equations for the populations of the upper hyper-fine ground state and the excited state $\left \langle \hat{\sigma}_1^{22}\right \rangle$, $\left \langle \hat{\sigma}_1^{33}\right \rangle$  take the form
\begin{align}
&\partial_{t}\langle\hat{\sigma}_1^{22}\rangle=\gamma\langle\hat{\sigma}_1^{33}\rangle-ig\left(\langle\hat{a}^{\dagger}\hat{\sigma}_1^{23}\rangle-\langle\hat{a}\hat{\sigma}_1^{32}\rangle\right)\nonumber \\
& + i\Omega_m(e^{i\omega_mt}\langle\hat{\sigma}_1^{12}\rangle-e^{-i\omega_mt}\langle\hat{\sigma}_1^{21}\rangle)\nonumber \\
& + \frac{dW}{dt}\sqrt{\eta\kappa/2}e^{i\omega_{p}t}\left(\langle\hat{a}\hat{\sigma}_1^{22}\rangle - \langle\hat{a}\rangle\langle\hat{\sigma}_1^{22}\rangle\right)\nonumber \\
& + \frac{dW}{dt}\sqrt{\eta\kappa/2}e^{-i\omega_{p}t}\left(\langle\hat{a}^{\dagger}\hat{\sigma}_1^{22}\rangle - \langle\hat{a}^{\dagger}\rangle\langle\hat{\sigma}_1^{22}\rangle\right),
\end{align}

\begin{align}
&\partial_{t}\langle\hat{\sigma}_1^{33}\rangle=-\gamma\langle\hat{\sigma}_1^{33}\rangle+ig\left(\langle\hat{a}^{\dagger}\hat{\sigma}_1^{23}\rangle-\langle\hat{a}\hat{\sigma}_1^{32}\rangle\right)\nonumber \\
& + \frac{dW}{dt}\sqrt{\eta\kappa/2}e^{i\omega_{p}t}\left(\langle\hat{a}\hat{\sigma}_1^{33}\rangle - \langle\hat{a}\rangle\langle\hat{\sigma}_1^{33}\rangle\right)\nonumber\\
& + \frac{dW}{dt}\sqrt{\eta\kappa/2}e^{-i\omega_{p}t}\left(\langle\hat{a}^{\dagger}\hat{\sigma}_1^{33}\rangle - \langle\hat{a}^{\dagger}\rangle\langle\hat{\sigma}_1^{33}\rangle\right).
\end{align}
Here, we do not consider directly the equations for the population of the lowest hyper-fine ground state $\langle\hat{\sigma}_1^{11}\rangle$ because of  $\langle\hat{\sigma}_1^{11}\rangle =  1- \langle\hat{\sigma}_1^{22}\rangle - \langle\hat{\sigma}_1^{33}\rangle$.

Now, we present the equations for the second-order mean quantities. The equation for the intra-cavity photon number reads
\begin{align}
&\partial_{t}\langle\hat{a}^{\dagger}\hat{a}\rangle=-\kappa\langle\hat{a}^{\dagger}\hat{a}\rangle+iNg\left(\langle\hat{a}^{\dagger}\hat{\sigma}_1^{23}\rangle+\langle\hat{a}\hat{\sigma}_1^{32}\rangle\right)\nonumber\\
& + i\Omega_{p}\sqrt{\kappa/2}(e^{i\omega_pt}\langle\hat{a}\rangle-e^{-i\omega_pt}\langle\hat{a}^{\dagger}\rangle)\nonumber\\
& + \frac{dW}{dt}\sqrt{\eta\kappa/2}(e^{i\omega_{p}t}\langle\hat{a}^{\dagger}\hat{a}\hat{a}\rangle
 + e^{-i\omega_{p}t}\langle\hat{a}\hat{a}^{\dagger}\hat{a}^{\dagger}\rangle). 
\end{align}
The  equation for the photon-photon correlation reads
\begin{align}
&\partial_{t}\langle\hat{a}\hat{a}\rangle=(-2i\omega_c-\kappa)\langle\hat{a}\hat{a}\rangle\nonumber \\
& +2iNg\langle\hat{a}\hat{\sigma}_1^{23}\rangle - 2i\Omega_{p}\sqrt{\kappa/2}e^{-i\omega_pt}\langle\hat{a}\rangle\nonumber \\
& + \frac{dW}{dt}\sqrt{\eta\kappa/2}(e^{i\omega_{p}t}\langle\hat{a}\hat{a}\hat{a}\rangle  + e^{-i\omega_{p}t}\langle\hat{a}^{\dagger}\hat{a}\hat{a}\rangle). 
\end{align}
The equations for the atom-photon correlations read
\begin{align}
& \partial_{t}\langle\hat{a}^{\dagger}\hat{\sigma}_1^{12}\rangle=\left(i\omega_c-i\omega_{21}-\kappa/2-\chi/4\right)\langle\hat{a}^{\dagger}\hat{\sigma}_1^{12}\rangle\nonumber \\
&  + ig[\left(N-1\right)\langle\hat{\sigma}_1^{12}\hat{\sigma}_2^{32}\rangle - \langle\hat{a}^{\dagger}\hat{a}^{\dagger}\hat{\sigma}_1^{13}\rangle]\nonumber \\
& + i\Omega_{p}\sqrt{\kappa/2}e^{i\omega_pt}\langle\hat{\sigma}_1^{12}\rangle \nonumber \\
&- i\Omega_me^{i\omega_mt}\left(\langle\hat{a}\rangle-\langle\hat{a}^{\dagger}\hat{\sigma}_1^{33}\rangle-2i\langle\hat{a}^{\dagger}\hat{\sigma}_1^{22}\rangle\right)\nonumber \\
& + \frac{dW}{dt}\sqrt{\eta\kappa/2}e^{i\omega_{p}t}(\langle\hat{a}^{\dagger}\hat{a}\hat{\sigma}_1^{12}\rangle - \langle\hat{a}\rangle\langle\hat{a}^{\dagger}\hat{\sigma}_1^{12}\rangle)\nonumber \\
& + \frac{dW}{dt}\sqrt{\eta\kappa/2}e^{-i\omega_{p}t}(\langle\hat{a}^{\dagger}\hat{a}^{\dagger}\hat{\sigma}_1^{12}\rangle - \langle\hat{a}^{\dagger}\rangle\langle\hat{a}^{\dagger}\hat{\sigma}_1^{12}\rangle),
\end{align}

\begin{align}
& \partial_{t}\langle\hat{a}^{\dagger}\hat{\sigma}_1^{13}\rangle=(i\omega_c-i\omega_{21}-i\omega_{32}-\kappa/2-\gamma/2\nonumber \\
& -\chi/4)\langle\hat{a}^{\dagger}\hat{\sigma}_1^{13}\rangle + ig[\left(N-1\right)\langle\hat{\sigma}_1^{32}\hat{\sigma}_2^{13}\rangle- \langle\hat{a}^{\dagger}\hat{a}\hat{\sigma}_1^{12}\rangle]\nonumber \\
& + i\Omega_{p}\sqrt{\kappa/2}e^{i\omega_pt}\langle\hat{\sigma}_1^{13}\rangle-i\Omega_me^{i\omega_mt}\langle\hat{a}^{\dagger}\hat{\sigma}_1^{21}\rangle\nonumber \\
& + \frac{dW}{dt}\sqrt{\eta\kappa/2}e^{i\omega_{p}t}(\langle\hat{a}^{\dagger}\hat{a}\hat{\sigma}_1^{13}\rangle - \langle\hat{a}\rangle\langle\hat{a}^{\dagger}\hat{\sigma}_1^{13}\rangle)\nonumber \\
& + \frac{dW}{dt}\sqrt{\eta\kappa/2}e^{-i\omega_{p}t}(\langle\hat{a}^{\dagger}\hat{a}^{\dagger}\hat{\sigma}_1^{13}\rangle - \langle\hat{a}^{\dagger}\rangle\langle\hat{a}^{\dagger}\hat{\sigma}_1^{13}\rangle),
\end{align}

\begin{align}
&\partial_{t}\langle\hat{a}^{\dagger}\hat{\sigma}_1^{23}\rangle=\left(i\omega_c-i\omega_{32}-\chi -\kappa/2-\gamma\right)\langle\hat{a}^{\dagger}\hat{\sigma}_1^{32}\rangle  \nonumber \\
&+ ig[\left(N-1\right)\langle\hat{\sigma}_1^{32}\hat{\sigma}_2^{23}\rangle + \langle\hat{a}^{\dagger}\hat{a}\hat{\sigma}_1^{33}\rangle- \langle\hat{a}^{\dagger}\hat{a}\hat{\sigma}_1^{22}\rangle]\nonumber \\
& + i\Omega_{p}\sqrt{\kappa/2}e^{i\omega_pt}\langle\hat{\sigma}_1^{23}\rangle+i\Omega_me^{i\omega_mt}\langle\hat{a}^{\dagger}\hat{\sigma}_1^{13}\rangle\nonumber \\
& + \frac{dW}{dt}\sqrt{\eta\kappa/2}e^{i\omega_{p}t}(\langle\hat{a}^{\dagger}\hat{a}\hat{\sigma}_1^{23}\rangle - \langle\hat{a}^{\dagger}\rangle\langle\hat{a}\hat{\sigma}_1^{23}\rangle)\nonumber \\
& + \frac{dW}{dt}\sqrt{\eta\kappa/2}e^{-i\omega_{p}t}(\langle\hat{a}^{\dagger}\hat{a}^{\dagger}\hat{\sigma}_1^{23}\rangle - \langle\hat{a}^{\dagger}\rangle\langle\hat{a}^{\dagger}\hat{\sigma}_1^{23}\rangle),
\end{align}

\begin{align}
&\partial_{t}\langle\hat{a}^{\dagger}\hat{\sigma}_1^{22}\rangle=(i\omega_c-\kappa/2+\gamma)\langle\hat{a}^{\dagger}\hat{\sigma}_1^{22}\rangle\nonumber \\
& + ig[\left(N-1\right)\langle\hat{\sigma}_1^{32}\hat{\sigma}_2^{23}\rangle+ \langle\hat{a}^{\dagger}\hat{a}\hat{\sigma}_1^{32}\rangle- \langle\hat{a}^{\dagger}\hat{a}^{\dagger}\hat{\sigma}_1^{23}\rangle]\nonumber \\
& + i\Omega_{p}\sqrt{\kappa/2}e^{i\omega_pt}\langle\hat{\sigma}_1^{22}\rangle  \nonumber \\
&+i\Omega_m (e^{i\omega_mt}\langle\hat{a}^{\dagger}\hat{\sigma}_1^{12}\rangle-e^{-i\omega_mt}\langle\hat{a}^{\dagger}\hat{\sigma}_1^{21}\rangle)\nonumber \\
& + \frac{dW}{dt}\sqrt{\eta\kappa/2}e^{i\omega_{p}t}(\langle\hat{a}\hat{\sigma}_1^{12}\hat{\sigma}_2^{12}\rangle - \langle\hat{a}\rangle\langle\hat{\sigma}_1^{12}\hat{\sigma}_2^{12}\rangle)\nonumber \\
& + \frac{dW}{dt}\sqrt{\eta\kappa/2}e^{-i\omega_{p}t}(\langle\hat{a}^{\dagger}\hat{a}^{\dagger}\hat{\sigma}_1^{22}\rangle - \langle\hat{a}^{\dagger}\rangle\langle\hat{a}^{\dagger}\hat{\sigma}_1^{22}\rangle),
\end{align}
\begin{align}
&\partial_{t}\langle\hat{a}^{\dagger}\hat{\sigma}_1^{33}\rangle=(i\omega_c-\kappa/2-\gamma)\langle\hat{a}^{\dagger}\hat{\sigma}_1^{33}\rangle \nonumber \\
&+ i\Omega_{p}\sqrt{\kappa/2}e^{i\omega_pt}\langle\hat{\sigma}_1^{33}\rangle \nonumber \\
& + ig[\left(N-1\right)\langle\hat{\sigma}_1^{33}\hat{\sigma}_2^{32}\rangle - \langle\hat{a}^{\dagger}\hat{a}\hat{\sigma}_1^{32}\rangle+ \langle\hat{a}^{\dagger}\hat{a}^{\dagger}\hat{\sigma}_1^{23}\rangle]\nonumber \\
& + \frac{dW}{dt}\sqrt{\eta\kappa/2}e^{i\omega_{p}t}(\langle\hat{a}^{\dagger}\hat{a}\hat{\sigma}_1^{33}\rangle - \langle\hat{a}\rangle\langle\hat{a}^{\dagger}\hat{\sigma}_1^{33}\rangle)\nonumber \\
& + \frac{dW}{dt}\sqrt{\eta\kappa/2}e^{-i\omega_{p}t}(\langle\hat{a}^{\dagger}\hat{a}^{\dagger}\hat{\sigma}_1^{33}\rangle - \langle\hat{a}^{\dagger}\rangle\langle\hat{a}^{\dagger}\hat{\sigma}_1^{33}\rangle),
\end{align}

\begin{align}
&\partial_{t}\langle\hat{a}\hat{\sigma}_1^{12}\rangle=(-i\omega_c-i\omega_{21}-\kappa/2 - \chi/4)\langle\hat{a}\hat{\sigma}_1^{12}\rangle \nonumber \\
&- ig[\left(N-1\right)\langle\hat{\sigma}_1^{12}\hat{\sigma}_2^{23}\rangle +\langle\hat{a}^{\dagger}\hat{a}\hat{\sigma}_1^{13}\rangle+\langle\hat{\sigma}_1^{13}\rangle]\nonumber \\
& - i\Omega_{p}\sqrt{\kappa/2}e^{i\omega_pt}\langle\hat{\sigma}_1^{12}\rangle \nonumber \\
&- i\Omega_me^{-i\omega_dt}(\langle\hat{a}\rangle-\langle\hat{a}\hat{\sigma}_1^{33}\rangle -\langle\hat{a}\hat{\sigma}_1^{22}\rangle)\nonumber \\
& + \frac{dW}{dt}\sqrt{\eta\kappa/2}e^{i\omega_{p}t}(\langle\hat{a}\hat{a}\hat{\sigma}_1^{12}\rangle - \langle\hat{a}\rangle\langle\hat{a}\hat{\sigma}_1^{12}\rangle)\nonumber \\
& + \frac{dW}{dt}\sqrt{\eta\kappa/2}e^{-i\omega_{p}t}(\langle\hat{a}\hat{a}^{\dagger}\hat{\sigma}_1^{12}\rangle - \langle\hat{a}^{\dagger}\rangle\langle\hat{a}\hat{\sigma}_1^{12}\rangle),
\end{align}

\begin{align}
& \partial_{t}\langle\hat{a}\hat{\sigma}_1^{13}\rangle=(-i\omega_c-i\omega_{21}-i\omega_{32}- \chi/4 -\kappa/2-\gamma/2)\langle\hat{a}\hat{\sigma}_1^{13}\rangle\nonumber \\
& - ig[\left(N-1\right)\langle\hat{\sigma}_1^{23}\hat{\sigma}_2^{13}\rangle+\langle\hat{a}\hat{a}\hat{\sigma}_1^{12}\rangle]\nonumber \\
& - i\Omega_{p}\sqrt{\kappa/2}e^{i\omega_pt}\langle\hat{\sigma}_1^{13}\rangle-i\Omega_me^{i\omega_dt}\langle\hat{a}\rangle\langle\hat{\sigma}_1^{23}\rangle\nonumber \\
& + \frac{dW}{dt}\sqrt{\eta\kappa/2}e^{i\omega_{p}t}(\langle \hat{a}\hat{a}\hat{\sigma}_1^{13}\rangle - \langle\hat{a}\rangle\langle\hat{a}\hat{\sigma}_1^{13}\rangle)\nonumber \\
& + \frac{dW}{dt}\sqrt{\eta\kappa/2}e^{-i\omega_{p}t}(\langle\hat{a}\hat{a}^{\dagger}\hat{\sigma}_1^{13}\rangle - \langle\hat{a}^{\dagger}\rangle\langle\hat{a}\hat{\sigma}_1^{13}\rangle),
\end{align}

\begin{align}
&\partial_{t}\langle\hat{a}\hat{\sigma}_1^{23}\rangle=(-i\omega_c+i\omega_{32}-\kappa/2-\gamma/2 - \chi)\langle\hat{a}\hat{\sigma}_1^{23}\rangle \nonumber \\
&- ig[\left(N-1\right)\langle\hat{\sigma}_1^{23}\hat{\sigma}_2^{23}\rangle-\langle\hat{a}\hat{a}\hat{\sigma}_1^{33}\rangle+\langle\hat{a}\hat{a}\hat{\sigma}_1^{22}\rangle]\nonumber \\
& - i\Omega_{p}\sqrt{\kappa/2}e^{i\omega_pt}\langle\hat{\sigma}_1^{23}\rangle\rangle+i\Omega_me^{i\omega_dt}\langle\hat{a}\rangle\langle\hat{\sigma}_1^{13}\rangle\nonumber \\
& + \frac{dW}{dt}\sqrt{\eta\kappa/2}e^{i\omega_{p}t}(\langle\hat{a}\hat{a}\hat{\sigma}_1^{23}\rangle - \langle\hat{a}\rangle\langle\hat{a}\hat{\sigma}_1^{23}\rangle)\nonumber \\
& + \frac{dW}{dt}\sqrt{\eta\kappa/2}e^{-i\omega_{p}t}(\langle\hat{a}\hat{a}^{\dagger}\hat{\sigma}_1^{23}\rangle - \langle\hat{a}^{\dagger}\rangle\langle\hat{a}\hat{\sigma}_1^{23}\rangle).
\end{align}

The equations for the atom-atom correlations read
\begin{align}
& \partial_{t}\langle\hat{\sigma}_1^{12}\hat{\sigma}_2^{12}\rangle = (-2i\omega_{21}-\chi/2)\langle\hat{\sigma}_1^{12}\hat{\sigma}_2^{12}\rangle  - 2ig\langle\hat{a}^{\dagger}\hat{\sigma}_1^{12}\hat{\sigma}_2^{13}\rangle\nonumber \\
& + 2i\Omega_me^{-i\omega_mt}\left(\langle\hat{\sigma}_1^{33}\hat{\sigma}_2^{12}\rangle+2\langle\hat{\sigma}_1^{22}\hat{\sigma}_2^{12}\rangle-\langle\hat{\sigma}_1^{12}\rangle\right)\nonumber \\
& + \frac{dW}{dt}\sqrt{\eta\kappa/2}e^{i\omega_{p}t}\left(\langle\hat{a}\hat{\sigma}_1^{12}\hat{\sigma}_2^{12}\rangle - \langle\hat{a}\rangle\langle\hat{\sigma}_1^{12}\hat{\sigma}_2^{12}\rangle\right)\nonumber \\
& + \frac{dW}{dt}\sqrt{\eta\kappa/2}e^{-i\omega_{p}t}\left(\langle\hat{a}^{\dagger}\hat{\sigma}_1^{12}\hat{\sigma}_2^{12}\rangle - \langle\hat{a}^{\dagger}\rangle\langle\hat{\sigma}_1^{12}\hat{\sigma}_2^{12}\rangle\right),
\end{align}
\begin{align}
&\partial_{t}\langle\hat{\sigma}_1^{22}\hat{\sigma}_2^{22}\rangle=2\gamma\langle\hat{\sigma}_1^{22}\hat{\sigma}_2^{33}\rangle-2ig(\langle\hat{a}^{\dagger}\hat{\sigma}_1^{22}\hat{\sigma}_2^{23}\rangle - \langle\hat{a}\hat{\sigma}_1^{22}\hat{\sigma}_2^{32}\rangle)\nonumber \\
& + 2i\Omega_m(e^{-i\omega_mt}\langle\hat{\sigma}_1^{22}\hat{\sigma}_2^{12}\rangle-e^{i\omega_mt}\langle\hat{\sigma}_1^{22}\hat{\sigma}_2^{21}\rangle)\nonumber \\
& + \frac{dW}{dt}\sqrt{\eta\kappa/2}e^{i\omega_{p}t}\left(\langle\hat{a}\hat{\sigma}_1^{22}\hat{\sigma}_2^{22}\rangle - \langle\hat{a}\rangle\langle\hat{\sigma}_1^{22}\hat{\sigma}_2^{22}\rangle\right)\nonumber \\
& + \frac{dW}{dt}\sqrt{\eta\kappa/2}e^{-i\omega_{p}t}\left(\langle\hat{a}^{\dagger}\hat{\sigma}_1^{22}\hat{\sigma}_2^{22}\rangle - \langle\hat{a}^{\dagger}\rangle\langle\hat{\sigma}_1^{22}\hat{\sigma}_2^{22}\rangle\right),
\end{align}
\begin{align}
&\partial_{t}\langle\hat{\sigma}_1^{23}\hat{\sigma}_2^{23}\rangle=(-2i\omega_{32}-\gamma-2\chi)\langle\hat{\sigma}_1^{23}\hat{\sigma}_2^{23}\rangle \nonumber \\
&+ 2ig(\langle\hat{a}\hat{\sigma}_1^{33}\hat{\sigma}_2^{23}\rangle- \langle\hat{a}\hat{\sigma}_1^{22}\hat{\sigma}_2^{23}\rangle) + 2i\Omega_me^{i\omega_mt}\langle\hat{\sigma}_1^{23}\hat{\sigma}_2^{13}\rangle\nonumber \\
& + \frac{dW}{dt}\sqrt{\eta\kappa/2}e^{i\omega_{p}t}\left(\langle\hat{a}\hat{\sigma}_1^{23}\hat{\sigma}_2^{23}\rangle - \langle\hat{a}\rangle\langle\hat{\sigma}_1^{23}\hat{\sigma}_2^{23}\rangle\right)\nonumber \\
& + \frac{dW}{dt}\sqrt{\eta\kappa/2}e^{-i\omega_{p}t}\left(\langle\hat{a}^{\dagger}\hat{\sigma}_1^{23}\hat{\sigma}_2^{23}\rangle - \langle\hat{a}^{\dagger}\rangle\langle\hat{\sigma}_1^{23}\hat{\sigma}_2^{23}\rangle\right),
\end{align}
\begin{align}
&\partial_{t}\langle\hat{\sigma}_1^{33}\hat{\sigma}_2^{33}\rangle=-2\gamma\langle\hat{\sigma}_1^{33}\hat{\sigma}_2^{33}\rangle\nonumber \\
&+2ig(\langle\hat{a}^{\dagger}\hat{\sigma}_1^{33}\hat{\sigma}_2^{23}\rangle - \langle\hat{a}\hat{\sigma}_1^{33}\hat{\sigma}_2^{32}\rangle)\nonumber \\
& + \frac{dW}{dt}\sqrt{\eta\kappa/2}e^{i\omega_{p}t}\left(\langle\hat{a}\hat{\sigma}_1^{33}\hat{\sigma}_2^{33}\rangle - \langle\hat{a}\rangle\langle\hat{\sigma}_1^{33}\hat{\sigma}_2^{33}\rangle\right)\nonumber \\
& + \frac{dW}{dt}\sqrt{\eta\kappa/2}e^{-i\omega_{p}t}\left(\langle\hat{a}^{\dagger}\hat{\sigma}_1^{33}\hat{\sigma}_2^{33}\rangle - \langle\hat{a}^{\dagger}\rangle\langle\hat{\sigma}_1^{33}\hat{\sigma}_2^{33}\rangle\right),
\end{align}
\begin{align}
&\partial_{t}\langle\hat{\sigma}_1^{13}\hat{\sigma}_2^{13}\rangle=(-2i\omega_{21}+2i\omega_{32}-\gamma -\chi/2)\langle\hat{\sigma}_1^{13}\hat{\sigma}_2^{13}\rangle \nonumber \\
& - 2ig\langle\hat{a}\hat{\sigma}_1^{12}\hat{\sigma}_2^{13}\rangle+2i\Omega_me^{i\omega_mt}\langle\hat{\sigma}_1^{23}\hat{\sigma}_2^{13}\rangle\nonumber \\
& + \frac{dW}{dt}\sqrt{\eta\kappa/2}e^{i\omega_{p}t}\left(\langle\hat{a}\hat{\sigma}_1^{13}\hat{\sigma}_2^{13}\rangle - \langle\hat{a}\rangle\langle\hat{\sigma}_1^{13}\hat{\sigma}_2^{13}\rangle\right)\nonumber \\
& + \frac{dW}{dt}\sqrt{\eta\kappa/2}e^{-i\omega_{p}t}\left(\langle\hat{a}^{\dagger}\hat{\sigma}_1^{13}\hat{\sigma}_2^{13}\rangle - \langle\hat{a}^{\dagger}\rangle\langle\hat{\sigma}_1^{13}\hat{\sigma}_2^{13}\rangle\right),
\end{align}

\begin{align}
&\partial_{t}\langle\hat{\sigma}_1^{12}\hat{\sigma}_2^{21}\rangle=-\chi/2\langle\hat{\sigma}_1^{12}\hat{\sigma}_2^{21}\rangle \nonumber \\
&-ig(\langle\hat{a}^{\dagger}\hat{\sigma}_1^{21}\hat{\sigma}_2^{13}\rangle-\langle\hat{a}\hat{\sigma}_1^{12}\hat{\sigma}_2^{31}\rangle)\nonumber \\
& + \frac{dW}{dt}\sqrt{\eta\kappa/2}e^{i\omega_{p}t}(\langle\hat{a}\hat{\sigma}_1^{21}\hat{\sigma}_1^{12}\rangle - \langle\hat{\sigma}_1^{21}\rangle\langle\hat{a}\hat{\sigma}_1^{12}\rangle)\nonumber \\
& + \frac{dW}{dt}\sqrt{\eta\kappa/2}e^{-i\omega_{p}t}(\langle\hat{a}^{\dagger}\hat{\sigma}_1^{21}\hat{\sigma}_1^{12}\rangle - \langle\hat{a}^{\dagger}\rangle\langle \hat{\sigma}_1^{21}\hat{\sigma}_1^{12}\rangle),
\end{align}

\begin{align}
&\partial_{t}\langle\hat{\sigma}_1^{12}\hat{\sigma}_2^{13}\rangle=(-2i\omega_{21}-i\omega_{32}-\gamma/2- \chi/2)\langle\hat{\sigma}_1^{12}\hat{\sigma}_2^{13}\rangle\nonumber \\
&- ig(\langle\hat{a}^{\dagger}\hat{\sigma}_1^{13}\hat{\sigma}_2^{13}\rangle+\langle\hat{a}\hat{\sigma}_1^{12}\hat{\sigma}_2^{12}\rangle)\nonumber \\
& - i\Omega_me^{-i\omega_mt}\left(\langle\hat{\sigma}_1^{13}\rangle-\langle\hat{\sigma}_1^{33}\hat{\sigma}_2^{13}\rangle-\langle\hat{\sigma}_1^{13}\hat{\sigma}_2^{23}\rangle-2\langle\hat{\sigma}_1^{22}\hat{\sigma}_2^{13}\rangle\right)\nonumber \\
& + \frac{dW}{dt}\sqrt{\eta\kappa/2}e^{i\omega_{p}t}(\langle\hat{\sigma}_1^{13}\hat{a}\hat{\sigma}_1^{12}\rangle- \langle\hat{a}\rangle\langle\hat{\sigma}_1^{13}\hat{\sigma}_1^{12}\rangle)\nonumber \\
& + \frac{dW}{dt}\sqrt{\eta\kappa/2}e^{-i\omega_{p}t}(\langle\hat{\sigma}_1^{13}\hat{a}^{\dagger}\hat{\sigma}_1^{12}\rangle - \langle\hat{a}^{\dagger}\rangle\langle \hat{\sigma}_1^{13}\hat{\sigma}_1^{12}\rangle),
\end{align}
\begin{align}
&\partial_{t}\langle\hat{\sigma}_1^{21}\hat{\sigma}_2^{13}\rangle=(-i\omega_{32}-\gamma/2- \chi/2)\langle\hat{\sigma}_1^{21}\hat{\sigma}_2^{13}\rangle\nonumber  \\
& + ig(\langle\hat{a}\hat{\sigma}_1^{31}\hat{\sigma}_2^{13}\rangle - \langle\hat{a}\hat{\sigma}_1^{21}\hat{\sigma}_2^{12}\rangle) - i\Omega_me^{-i\omega_mt}\langle\hat{\sigma}_1^{21}\hat{\sigma}_2^{23}\rangle \nonumber \\
& + i\Omega_me^{i\omega_mt}\left(\langle\hat{\sigma}_1^{13}\rangle-\langle\hat{\sigma}_1^{33}\hat{\sigma}_2^{13}\rangle-2\langle\hat{\sigma}_1^{22}\hat{\sigma}_2^{13}\rangle\right)\nonumber \\
& + \frac{dW}{dt}\sqrt{\eta\kappa/2}e^{i\omega_{p}t}(\langle\hat{\sigma}_1^{13}\hat{a}\hat{\sigma}_1^{21}\rangle - \langle\hat{a}\rangle\langle \hat{\sigma}_1^{13}\hat{\sigma}_1^{21})\nonumber \\
& + \frac{dW}{dt}\sqrt{\eta\kappa/2}e^{-i\omega_{p}t}(\langle\hat{\sigma}_1^{13}\hat{a}^{\dagger}\hat{\sigma}_1^{21}\rangle - \langle\hat{a}^{\dagger}\rangle\langle \hat{\sigma}_1^{13}\hat{\sigma}_1^{21}\rangle),
\end{align}

\begin{align}
&\partial_{t}\langle\hat{\sigma}_1^{32}\hat{\sigma}_2^{23}\rangle=(-\gamma- 2\chi)\langle\hat{\sigma}_1^{32}\hat{\sigma}_2^{23}\rangle\nonumber \\
& - ig(\langle\hat{a}^{\dagger}\hat{\sigma}_1^{33}\hat{\sigma}_2^{23}\rangle-\langle\hat{a}^{\dagger}\hat{\sigma}_1^{22}\hat{\sigma}_2^{23}\rangle - \langle\hat{a}\hat{\sigma}_1^{33}\hat{\sigma}_2^{32}\rangle+\langle\hat{a}\hat{\sigma}_1^{22}\hat{\sigma}_2^{32}\rangle)\nonumber \\
& + i\Omega_m(e^{i\omega_mt}\langle\hat{\sigma}_1^{32}\hat{\sigma}_2^{13}\rangle-e^{-i\omega_mt}\langle\hat{\sigma}_1^{23}\hat{\sigma}_2^{31}\rangle)\nonumber \\
& + \frac{dW}{dt}\sqrt{\eta\kappa/2}e^{i\omega_{p}t}(\langle\hat{\sigma}_1^{23}\hat{a}\hat{\sigma}_1^{32}\rangle - 2\langle\hat{a}\rangle\langle \hat{\sigma}_1^{23}\hat{\sigma}_1^{32}\rangle)\nonumber \\
& + \frac{dW}{dt}\sqrt{\eta\kappa/2}e^{-i\omega_{p}t}(\langle\hat{\sigma}_1^{23}\hat{a}^{\dagger}\hat{\sigma}_1^{32}\rangle - \langle\hat{a}^{\dagger}\rangle\langle \hat{\sigma}_1^{23}\hat{\sigma}_1^{32}\rangle),
\end{align}
\begin{align}
&\partial_{t}\langle\hat{\sigma}_1^{32}\hat{\sigma}_2^{13}\rangle=(-i\omega_{21}-\gamma- 5/4\chi)\langle\hat{\sigma}_1^{32}\hat{\sigma}_2^{13}\rangle \nonumber \\
&+ ig(\langle\hat{a}^{\dagger}\hat{\sigma}_1^{22}\hat{\sigma}_2^{13}\rangle - \langle\hat{a}^{\dagger}\hat{\sigma}_1^{33}\hat{\sigma}_2^{13}\rangle- \langle\hat{a}\hat{\sigma}_1^{12}\hat{\sigma}_2^{32}\rangle)\nonumber \\
& -i\Omega_me^{-i\omega_mt}\left(\langle\hat{\sigma}_1^{31}\hat{\sigma}_2^{13}\rangle-\langle\hat{\sigma}_1^{32}\hat{\sigma}_2^{23}\rangle\right)\nonumber \\
& + \frac{dW}{dt}\sqrt{\eta\kappa/2}e^{i\omega_{p}t}(\langle\hat{\sigma}_1^{13}\hat{a}\hat{\sigma}_1^{32}\rangle - \langle\hat{a}\rangle\langle \hat{\sigma}_1^{13}\hat{\sigma}_1^{32}\rangle)\nonumber \\
& + \frac{dW}{dt}\sqrt{\eta\kappa/2}e^{-i\omega_{p}t}(\langle\hat{\sigma}_1^{13}\hat{a}^{\dagger}\hat{\sigma}_1^{32}\rangle - \langle\hat{a}^{\dagger}\rangle\langle \hat{\sigma}_1^{13}\hat{\sigma}_1^{32}\rangle),
\end{align}
\begin{align}
&\partial_{t}\langle\hat{\sigma}_1^{12}\hat{\sigma}_2^{32}\rangle=(-i\omega_{21}+i\omega_{32}-\gamma/2- 5/4\chi)\langle\hat{\sigma}_1^{12}\hat{\sigma}_2^{32}\rangle \nonumber \\
& + ig(\langle\hat{a}^{\dagger}\hat{\sigma}_1^{22}\hat{\sigma}_2^{12}\rangle - \langle\hat{a}^{\dagger}\hat{\sigma}_1^{33}\hat{\sigma}_2^{12}\rangle - \langle\hat{a}^{\dagger}\hat{\sigma}_1^{32}\hat{\sigma}_2^{13}\rangle)\nonumber \\
& - i\Omega_me^{-i\omega_mt}\left(\langle\hat{\sigma}_1^{32}\rangle+\langle\hat{\sigma}_1^{12}\hat{\sigma}_2^{31}\rangle-\langle\hat{\sigma}_1^{33}\hat{\sigma}_2^{32}\rangle-\langle\hat{\sigma}_1^{22}\hat{\sigma}_2^{32}\rangle\right)\nonumber \\
& + \frac{dW}{dt}\sqrt{\eta\kappa/2}e^{i\omega_{p}t}(\langle\hat{\sigma}_1^{32}\hat{a}\hat{\sigma}_1^{12}\rangle - \langle\hat{a}\rangle\langle \hat{\sigma}_1^{32}\hat{\sigma}_1^{12}\rangle)\nonumber \\
& + \frac{dW}{dt}\sqrt{\eta\kappa/2}e^{-i\omega_{p}t}(\langle\hat{\sigma}_1^{32}\hat{a}^{\dagger}\hat{\sigma}_1^{12}\rangle - \langle\hat{a}^{\dagger}\rangle\langle \hat{\sigma}_1^{32}\hat{\sigma}_1^{12}\rangle),
\end{align}
\begin{align}
&\partial_{t}\langle\hat{\sigma}_1^{12}\hat{\sigma}_2^{23}\rangle=(-i\omega_{21}-i\omega_{32}-\gamma/2 - 5/4\chi)\langle\hat{\sigma}_1^{12}\hat{\sigma}_2^{23}\rangle\nonumber \\
&- ig(\langle a^{\dagger}\sigma_{1}^{23}\sigma_{2}^{13}\rangle+\langle a\sigma_{1}^{22}\sigma_{2}^{12}\rangle  - \langle a\sigma_{1}^{33}\sigma_{2}^{12}\rangle)\nonumber \\
& + i\Omega_{m}[e^{i\omega_{m}t}\langle\sigma_{1}^{12}\sigma_{2}^{13}\rangle\nonumber \\
&-e^{-i\omega_{m}t}\left(\langle\hat{\sigma}_1^{13}\rangle-\langle\hat{\sigma}_1^{33}\hat{\sigma}_2^{23}\rangle-\langle\hat{\sigma}_1^{22}\hat{\sigma}_2^{23}\rangle\right)]\nonumber \\
& + \frac{dW}{dt}\sqrt{\eta\kappa/2}e^{i\omega_{p}t}(\langle\hat{\sigma}_1^{23}\hat{a}\hat{\sigma}_1^{12}\rangle - \langle\hat{a}\rangle\langle \hat{\sigma}_1^{23}\hat{\sigma}_1^{12}\rangle)\nonumber \\
& + \frac{dW}{dt}\sqrt{\eta\kappa/2}e^{-i\omega_{p}t}(\langle\hat{\sigma}_1^{23}\hat{a}^{\dagger}\hat{\sigma}_1^{12}\rangle - \langle\hat{a}^{\dagger}\rangle\langle \hat{\sigma}_1^{23}\hat{\sigma}_1^{12}\rangle),\nonumber \\
\end{align}
\begin{align}
&\partial_{t}\langle\hat{\sigma}_1^{31}\hat{\sigma}_2^{13}\rangle=(-\gamma-\chi/2)\langle\hat{\sigma}_1^{31}\hat{\sigma}_2^{13}\rangle\nonumber \\
& + ig(\langle a^{\dagger}\sigma_{1}^{21}\sigma_{2}^{13}\rangle- \langle a\sigma_{1}^{12}\sigma_{2}^{31}\rangle) \nonumber \\
& - i\Omega_{m}[e^{i\omega_{m}t}\langle\sigma_{1}^{32}\sigma_{2}^{13}\rangle-e^{-i\omega_{m}t}\langle\sigma_{1}^{23}\sigma_{2}^{31}\rangle]\nonumber \\
& + \frac{dW}{dt}\sqrt{\eta\kappa/2}e^{i\omega_{p}t}(\langle\hat{a}\hat{\sigma}_1^{13}\hat{\sigma}_1^{31}\rangle - \langle\hat{a}\rangle\langle \hat{\sigma}_1^{13}\hat{\sigma}_1^{31}\rangle)\nonumber \\
& + \frac{dW}{dt}\sqrt{\eta\kappa/2}e^{-i\omega_{p}t}(\langle\hat{a}^{\dagger}\hat{\sigma}_1^{13}\hat{\sigma}_1^{31}\rangle - \langle\hat{a}^{\dagger}\rangle\langle \hat{\sigma}_1^{13}\hat{\sigma}_1^{31}\rangle),\nonumber \\
\end{align}

\begin{align}
& \partial_{t}\langle\hat{\sigma}_1^{23}\hat{\sigma}_2^{13}\rangle=(-i\omega_{21}-2i\omega_{32}-\gamma- 5/4\chi)\langle\hat{\sigma}_1^{23}\hat{\sigma}_2^{13}\rangle \nonumber \\
&- ig(\langle\hat{a}\sigma_{1}^{22}\sigma_{2}^{13}\rangle+\langle\hat{a}\sigma_{1}^{12}\sigma_{2}^{23}\rangle-\langle\hat{a}\hat{\sigma}_1^{33}\hat{\sigma}_2^{13}\rangle)\nonumber \\
& + i\Omega_{m}[e^{i\omega_{m}t}\langle\sigma_{1}^{13}\sigma_{2}^{13}\rangle+e^{-i\omega_{m}t}\langle\sigma_{1}^{23}\sigma_{2}^{23}\rangle]\nonumber \\
& + \frac{dW}{dt}\sqrt{\eta\kappa/2}e^{i\omega_{p}t}(\langle\hat{a}\hat{\sigma}_1^{13}\hat{\sigma}_1^{23}\rangle - \langle\hat{a}\rangle\langle \hat{\sigma}_1^{13}\hat{\sigma}_1^{23}\rangle)\nonumber \\
& + \frac{dW}{dt}\sqrt{\eta\kappa/2}e^{-i\omega_{p}t}(\langle\hat{a}^{\dagger}\hat{\sigma}_1^{13}\hat{\sigma}_1^{23}\rangle - \langle\hat{a}^{\dagger}\rangle\langle \hat{\sigma}_1^{13}\hat{\sigma}_1^{23}\rangle),
\end{align}

\begin{align}
&\partial_{t}\langle\hat{\sigma}_1^{22}\hat{\sigma}_2^{13}\rangle=(-i\omega_{21}-i\omega_{32}-\gamma/2-\chi/4)\langle\hat{\sigma}_1^{22}\hat{\sigma}_2^{13}\rangle \nonumber \\
& +\gamma_{32}\langle\hat{\sigma}_1^{33}\hat{\sigma}_2^{13}\rangle - ig(\langle a^{\dagger}\sigma_{1}^{23}\sigma_{2}^{13}\rangle + \langle a\sigma_{1}^{22}\sigma_{2}^{12}\rangle- \langle a\sigma_{1}^{32}\sigma_{2}^{13}\rangle) \nonumber \\
& + i\Omega_{m}[e^{i\omega_{m}t}\langle\sigma_{1}^{12}\sigma_{2}^{13}\rangle -e ^{-i\omega_{m}t}\left(\langle\sigma_{1}^{21}\sigma_{2}^{13}\rangle-\langle\sigma_{1}^{22}\sigma_{2}^{23}\right\rangle)]\nonumber \\
& + \frac{dW}{dt}\sqrt{\eta\kappa/2}e^{i\omega_{p}t}(\langle\hat{a}\hat{\sigma}_1^{13}\hat{\sigma}_1^{22}\rangle - \langle\hat{a}\rangle\langle \hat{\sigma}_1^{13}\hat{\sigma}_1^{22}\rangle)\nonumber \\
& + \frac{dW}{dt}\sqrt{\eta\kappa/2}e^{-i\omega_{p}t}(\langle\hat{a}^{\dagger}\hat{\sigma}_1^{13}\hat{\sigma}_1^{22}\rangle - \langle\hat{a}^{\dagger}\rangle\langle \hat{\sigma}_1^{13}\hat{\sigma}_1^{22}\rangle),
\end{align}

\begin{align}
&\partial_{t}\langle\hat{\sigma}_1^{33}\hat{\sigma}_2^{13}\rangle=(-i\omega_{21}-i\omega_{32}-3\gamma/2- \chi/4)\langle\hat{\sigma}_1^{33}\hat{\sigma}_2^{13}\rangle \nonumber \\
&- ig(\langle\hat{a}\sigma_{1}^{33}\sigma_{2}^{12}\rangle+\langle\hat{a}\sigma_{1}^{32}\sigma_{2}^{13}\rangle-\langle a^{\dagger}\sigma_{1}^{23}\sigma_{2}^{13}\rangle) \\
&+i\Omega_{m}e^{-i\omega_{m}t}\langle\sigma_{1}^{33}\sigma_{2}^{23}\rangle\nonumber \\
& + \frac{dW}{dt}\sqrt{\eta\kappa/2}e^{i\omega_{p}t}(\langle\hat{a}\hat{\sigma}_1^{13}\hat{\sigma}_1^{33}\rangle - \langle\hat{a}\rangle\langle \hat{\sigma}_1^{13}\hat{\sigma}_1^{33}\rangle)\nonumber \\
& + \frac{dW}{dt}\sqrt{\eta\kappa/2}e^{-i\omega_{p}t}(\langle\hat{a}^{\dagger}\hat{\sigma}_1^{13}\hat{\sigma}_1^{33}\rangle - \langle\hat{a}^{\dagger}\rangle\langle \hat{\sigma}_1^{13}\hat{\sigma}_1^{33}\rangle),
\end{align}

\begin{align}
&\partial_{t}\langle\hat{\sigma}_1^{33}\hat{\sigma}_2^{32}\rangle=(i\omega_{32}-3\gamma/2- \chi)\langle\hat{\sigma}_1^{33}\hat{\sigma}_2^{32}\rangle \nonumber \\
&- i\Omega_{m}e^{-i\omega_{m}t}\langle\sigma_{1}^{33}\sigma_{2}^{31}\rangle \nonumber \\
&+ ig(\langle\hat{a}^{\dagger}\sigma_{1}^{32}\sigma_{2}^{23}\rangle+\langle\hat{a}^{\dagger}\sigma_{1}^{22}\sigma_{2}^{33}\rangle  - \langle a^{\dagger}\sigma_{1}^{33}\hat{\sigma}_2^{33}\rangle-\langle\hat{a}\hat{\sigma}_1^{32}\hat{\sigma}_2^{32}\rangle)\nonumber \\
& + \frac{dW}{dt}\sqrt{\eta\kappa/2}e^{i\omega_{p}t}(\langle\hat{a}\hat{\sigma}_1^{32}\hat{\sigma}_1^{33}\rangle - \langle\hat{a}\rangle\langle \hat{\sigma}_1^{32}\hat{\sigma}_1^{33}\rangle)\nonumber \\
& + \frac{dW}{dt}\sqrt{\eta\kappa/2}e^{-i\omega_{p}t}(\langle\hat{a}^{\dagger}\hat{\sigma}_1^{32}\hat{\sigma}_1^{33}\rangle - \langle\hat{a}^{\dagger}\rangle\langle \hat{\sigma}_1^{32}\hat{\sigma}_1^{33}\rangle),
\end{align}

\begin{align}
&\partial_{t}\langle\hat{\sigma}_1^{22}\hat{\sigma}_2^{33}\rangle=-\gamma\langle\hat{\sigma}_1^{22}\hat{\sigma}_2^{33}\rangle+\gamma_{32}\langle\hat{\sigma}_1^{33}\hat{\sigma}_2^{33}\rangle\nonumber \\
&-ig(\langle a^{\dagger}\sigma_{1}^{33}\sigma_{2}^{23}\rangle- \langle a^{\dagger}\sigma_{1}^{22}\sigma_{2}^{23}\rangle\nonumber  - \langle a\sigma_{1}^{33}\sigma_{2}^{32}\rangle+\langle a\sigma_{1}^{22}\sigma_{2}^{32}\rangle)\nonumber \\
& +i\Omega_{m}(e^{i\omega_{m}t}\langle\sigma_{1}^{33}\sigma_{2}^{12}\rangle- e^{-i\omega_{m}t}\langle\sigma_{1}^{33}\sigma_{2}^{21}\rangle)\nonumber \\
& + \frac{dW}{dt}\sqrt{\eta\kappa/2}e^{i\omega_{p}t}(\langle\hat{a}\hat{\sigma}_1^{33}\hat{\sigma}_1^{22}\rangle - \langle\hat{a}\rangle\langle \hat{\sigma}_1^{33}\hat{\sigma}_1^{22}\rangle)\nonumber \\
& + \frac{dW}{dt}\sqrt{\eta\kappa/2}e^{-i\omega_{p}t}(\langle\hat{a}^{\dagger}\hat{\sigma}_1^{33}\hat{\sigma}_1^{22}\rangle - \langle\hat{a}^{\dagger}\rangle\langle \hat{\sigma}_1^{33}\hat{\sigma}_1^{22}\rangle),
\end{align}

\begin{align}
&\partial_{t}\langle\hat{\sigma}_1^{22}\hat{\sigma}_2^{23}\rangle=(-i\omega_{32}-\gamma/2 - \chi)\langle\hat{\sigma}_1^{22}\hat{\sigma}_2^{23}\rangle+\gamma_{32}\langle\hat{\sigma}_1^{33}\hat{\sigma}_2^{23}\rangle \nonumber \\
&+ ig(\langle\hat{a}\sigma_{1}^{32}\sigma_{2}^{23}\rangle + \langle\hat{a}\sigma_{1}^{22}\sigma_{2}^{33}\rangle -\langle a^{\dagger}\sigma_{1}^{23}\sigma_{2}^{23}\rangle-\langle a\sigma_{1}^{22}\sigma_{2}^{22}\rangle)\nonumber \\
& + i\Omega_m[e^{i\omega_{m}t}(\langle\sigma_{1}^{12}\sigma_{2}^{23}\rangle+\langle\sigma_{1}^{22}\sigma_{2}^{13}\rangle)-e^{-i\omega_{m}t}\langle\sigma_{1}^{21}\sigma_{2}^{23}\rangle]\nonumber \\
& + \frac{dW}{dt}\sqrt{\eta\kappa/2}e^{i\omega_{p}t}(\langle\hat{a}\hat{\sigma}_1^{23}\hat{\sigma}_1^{22}\rangle - \langle\hat{a}\rangle\langle \hat{\sigma}_1^{23}\hat{\sigma}_1^{22}\rangle)\nonumber \\
& + \frac{dW}{dt}\sqrt{\eta\kappa/2}e^{-i\omega_{p}t}(\langle\hat{a}^{\dagger}\hat{\sigma}_1^{23}\hat{\sigma}_1^{22} - \langle\hat{a}^{\dagger}\rangle\langle \hat{\sigma}_1^{23}\hat{\sigma}_1^{22}\rangle),
\end{align}

\begin{align}
&\partial_{t}\langle\hat{\sigma}_1^{22}\hat{\sigma}_2^{12}\rangle=(-i\omega_{21}-\chi/4)\langle\hat{\sigma}_1^{22}\hat{\sigma}_2^{12}\rangle+\gamma\langle\hat{\sigma}_1^{33}\hat{\sigma}_2^{12}\rangle  \nonumber \\
&- ig(\langle\hat{a}^{\dagger}\sigma_{1}^{22}\sigma_{2}^{13}\rangle+\langle\hat{a}^{\dagger}\sigma_{1}^{12}\sigma_{2}^{23}\rangle - \langle a\sigma_{1}^{12}\sigma_{2}^{32}\rangle)\nonumber \\
& +i\Omega_{m}[e^{i\omega_{m}t}\langle\sigma_{1}^{12}\sigma_{2}^{12}\rangle-e^{-i\omega_{m}t}(\langle\sigma_{1}^{22}\rangle \nonumber\\
& -2\langle\sigma_{1}^{22}\sigma_{2}^{22}\rangle-\langle\sigma_{1}^{22}\sigma_{2}^{33}\rangle +\langle\sigma_{1}^{21}\sigma_{2}^{12}\rangle)]\nonumber \\
& + \frac{dW}{dt}\sqrt{\eta\kappa/2}e^{i\omega_{p}t}(\langle\hat{a}\hat{\sigma}_1^{12}\hat{\sigma}_1^{22} - \langle\hat{a}\rangle\langle \hat{\sigma}_1^{12}\hat{\sigma}_1^{22}\rangle)\nonumber \\
& + \frac{dW}{dt}\sqrt{\eta\kappa/2}e^{-i\omega_{p}t}(\langle\hat{a}^{\dagger}\hat{\sigma}_1^{12}\hat{\sigma}_1^{22}\rangle - \langle\hat{a}^{\dagger}\rangle\langle \hat{\sigma}_1^{12}\hat{\sigma}_1^{22}\rangle),
\end{align}

\begin{align}
&\partial_{t}\langle\hat{\sigma}_1^{33}\hat{\sigma}_2^{12}\rangle=(-i\omega_{21}-\gamma- \chi/4)\langle\hat{\sigma}_1^{33}\hat{\sigma}_2^{12}\rangle \nonumber \\
& - ig(\langle a^{\dagger}\sigma_{1}^{33}\sigma_{2}^{13}\rangle+\langle a\sigma_{1}^{12}\sigma_{2}^{32}\rangle - \langle a^{\dagger}\sigma_{1}^{12}\sigma_{2}^{23}\rangle)\nonumber \\
& - i\Omega_{m}e^{-i\omega_{m}t}(\langle\sigma_{1}^{33}\rangle-\langle\sigma_{1}^{33}\sigma_{2}^{33}\rangle-2\langle\sigma_{1}^{22}\sigma_{2}^{33}\rangle)\nonumber \\
& + \frac{dW}{dt}\sqrt{\eta\kappa/2}e^{i\omega_{p}t}(\langle\hat{a}\hat{\sigma}_1^{12}\hat{\sigma}_1^{33}\rangle - \langle\hat{a}\rangle\langle \hat{\sigma}_1^{12}\hat{\sigma}_1^{33}\rangle)\nonumber \\
& + \frac{dW}{dt}\sqrt{\eta\kappa/2}e^{-i\omega_{p}t}(\langle\hat{a}^{\dagger}\hat{\sigma}_1^{12}\hat{\sigma}_1^{33}\rangle - \langle\hat{a}^{\dagger}\rangle\langle \hat{\sigma}_1^{12}\hat{\sigma}_1^{33}\rangle).
\end{align}
By analyzing the mean-field equations closely, we find that these equations depend on also the quantities $\langle\hat{\sigma}_1^{21}\hat{\sigma}_2^{21}\rangle$, $\langle\hat{\sigma}_1^{21}\hat{\sigma}_2^{12}\rangle$. Since these quantities are complex conjugations of some quantities mentioned above, e.g. $\langle\hat{\sigma}_1^{21}\rangle=\langle\hat{\sigma}_1^{12}\rangle^*$, we do not need to consider the equations for them.

\end{document}